%% file: main.tex
\documentclass[10pt]{iopart}
\input{packages_commands}
\usepackage{esint}
\usepackage{amsmath}

\hypersetup{
     colorlinks   = true,
     linkcolor    = blue,
     citecolor    = blue,
     urlcolor     = blue
}

\begin{document}
\title{Initial observations in X-point target divertor discharges on MAST-U}
\author{N. Lonigro$^{1,2}$, K. Verhaegh$^{3}$, J. Harrison$^{1}$, B. Lipschultz$^{2}$, C. Bowman$^{1}$, F. Federici$^{4}$, J. Flanagan$^{5}$, D. Greenhouse$^{2}$, D. Moulton$^{1}$, P. Ryan$^{1}$, R. Scannell$^{1}$, S. Silburn$^{1}$, T. Wijkamp$^{6}$, D. Brida$^{7}$, C. Theiler$^{8}$, the EUROfusion Tokamak Exploitation Team$^{9}$ and the MAST Upgrade Team$^{10}$}
\address{$^1$UKAEA, Culham Campus, Abingdon, Oxfordshire,OX14 3DB, United Kingdom}
\address{$^2$York Plasma Institute, University of York, United Kingdom  }
\address{$^3$Department of Applied Physics, Eindhoven University of Technology, Netherlands}
\address{$^4$Oak Ridge National Laboratory, Oak Ridge, TN 37831, USA} 
\address{$5$University of Liverpool, Liverpool, L69 7ZX, UK}
\address{$^6$DIFFER, Eindhoven, Netherlands}
\address{$^7$ Max Planck Institute for Plasma Physics, Garching, Germany}  
\address{$^8$ Ecole Polytechnique Federale de Lausanne (EPFL), Swiss Plasma Center (SPC), Switzerland}
\address{$^9$See the author list of E. Joffrin et al. 2024 Nucl. Fusion 64 112019}
\address{$^{10}$See the author list of J.R. Harrison et al 2024 Nucl. Fusion 64 112017}
\ead{\mailto{nicola.lonigro@ukaea.uk}}

\begin{abstract}
The first high-power ($\geq$ 3 MW) H-mode experiments using a double-null X-point-target (XPT) divertor configuration have been performed on MAST-U. The XPT geometry is obtained by combining a large strike point radius, similar to the Super-X divertor (SXD), with an additional X-point near the separatrix in the baffled outer divertor chambers and leads to additional exhaust benefits over the SXD. The broader electron density profile near the secondary X-point leads to additional plasma-neutral interactions, evidenced by a broader hydrogenic emission profile, and resulting in larger power and ion sinks. The increase in plasma-neutral interactions also leads to lower target electron temperatures and heat fluxes. These benefits appear to extend to transients, and preliminary evidence of improved ELM buffering in the XPT is presented. These results showcase how multiple alternative divertor configuration strategies can be combined to improve momentum, power, and particle losses, which may be required for the challenging exhaust conditions of future reactors.
\end{abstract}


\section{Introduction}
\begin{figure}[]\centering
\includegraphics[width= 0.75\textwidth,trim={0 0 0 0},clip]{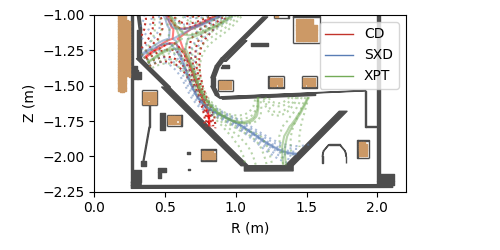}
\vspace{-0.1 cm}\caption{Comparison of experimental EFIT magnetic reconstructions of conventional (CD, \#49139),Super-X (SXD, \#49323), and X-point target (XPT, \#49320) divertor configurations.}
\label{fig:fig1_EFIT}
\vspace{-0.3cm} 
\end{figure}

Preventing damage to the internal surfaces of future fusion reactors will require handling the large particle and heat fluxes directed towards the divertor targets. Operating with a detached divertor will reduce the peak fluxes to the target through a combination of impurity radiation, radial transport, and plasma-neutral interactions, and is considered to be a requirement to ensure the peak heat fluxes and erosion rate are kept below material limits \cite{Wenninger_2014_DEMO_detach}. \\
However, conventional divertor strategies may still be insufficient for future devices due to a combination of the extreme conditions reached at the target, an insufficient resilience to changes in discharge conditions, and the requirement of large core radiation to reduce the power flowing to the divertor to manageable values. These challenges are exacerbated in compact designs such as STEP\cite{Henderson_2025_STEP} and ARC\cite{SPARC}. Thus, alternative divertor configurations (ADCs) are being studied on a variety of devices, such as MAST-U\cite{Harrison_2024}, TCV\cite{Theiler_2017}, AUG\cite{AUG_ADC}, and the planned DTT\cite{DTT}, to explore how the operational space of current and future devices can be expanded by using these divertor configurations. Some possible advantages include improved access to the detached regime, which makes finding an integrated core-edge solution easier, or deeper detachment states with lower peak particle and heat fluxes at the target for the same upstream conditions. Additional benefits can
include increased resilience of detachment to both fast transients \cite{HENDERSON_reattachment} and to slow changes in upstream parameters \cite{kool2024demonstrationsuperxdivertorexhaust}. \\
The Super-X divertor (SXD) is a promising ADC that moves the strike point to a large major radius, thus spreading the heat over a larger surface area and reducing the peak heat fluxes and target electron temperatures. Experiments on MAST-U in Super-X configuration, which combine the large radius effect with strong divertor baffling, have shown significant benefits over the conventional divertor. These include improved access to detachment\cite{verhaegh2024improved}, additional volumetric particle losses due to the increased volume available for plasma neutral interactions\cite{Verhaegh_2023_2}\cite{verhaegh2024nbi}, and additional power losses from plasma-molecule collisions\cite{osborne2024_NBI}, in agreement with SOLPS modeling \cite{moulton_super-x_2024}\cite{MAURIZIO2024101736}. \\
The ability of the divertor to buffer fast transients, such as Edge Localized Modes (ELMs), has also recently been investigated on multiple machines such as AUG \cite{ELM_asdex} and JET \cite{ELM_JET} due to the critical risk ELMs pose to the lifetime of the reactor components and the Super-X divertor on MAST-U has shown the possibility to buffer small ELMs without any extrinsic impurity seeding in conditions in which the conventional divertor configuration would be attached, as well as increasing buffering performance with increased divertor  neutral pressure\cite{Flanagan_2025}.\\
Another ADC showing good promise in modelling is the X-point target (XPT) configuration \cite{LaBombard_ADX}\cite{UMANSKY_XPT_ADX}\cite{Wigram_2019_XPT}, in which an additional X-point is created in the divertor chamber, near the separatrix. Previous results on the TCV tokamak showed no improvement in detachment access but a reduced sensitivity of the detachment front location near the 2nd null in this configuration \cite{Theiler_2017}. More recent work on the same device has demonstrated strong exhaust benefits by carefully positioning the secondary X-point very close to the separatrix. A significantly improved access to detachment has been observed in these single-null ohmic XPT discharges, in which the strike point is kept at the same major radius as the X-point, in addition to reduced peak particle and heat flux to the targets\cite{Lee_XPT}. In these experiments, a localization of the radiation near the secondary X-point is also observed, in agreement with the previous measurements of reduced sensitivity in its proximity \cite{Theiler_2017}, which can also be interpreted similarly to an X-point radiator \cite{Lee_XPT}.\\
In this work, MAST-U's unique capabilities are used to combine the effects of the Super-X divertor and the X-point target by creating the secondary X-points at a large major radius in the strongly baffled divertor chambers. This combination of strong baffling, a double-null XPT configuration, and a large strike point radius can further increase the divertor exhaust performance compared to the SXD and is the envisioned exhaust solution for the ARC reactor concept\cite{SPARC}. Example experimental magnetic reconstructions in conventional, Super-X, and X-point target configurations are compared in figure \ref{fig:fig1_EFIT}, where the SXD is shown to have a longer divertor leg and a larger strike point radius compared to the conventional divertor, while the XPT also has a secondary X-point in the divertor chamber and comparable strike point major radius to the SXD.\\
An overview of the XPT and SXD experimental discharges is given in section \ref{sec:XPT_discharge}. The results of the first H-mode experiments in double-null X-point target configuration on MAST-U are presented in section \ref{sec:XPT_exp} and compared against the SXD reference. These allow validating the previous results on the improved exhaust performance of the XPT obtained on TCV and include evidence of a wider plasma-neutral interaction region, resulting in additional power and momentum losses, and leading to lower target peak heat and particle fluxes. In section \ref{sec:discussion}, the results are compared against reduced models and the implications for future reactors are discussed. Finally, conclusions are given in section \ref{sec:conclusions} and the next steps are presented.
\section{XPT discharge overview}\label{sec:XPT_discharge}
\begin{figure}[]\centering
\begin{subfigure}[b]{0.34\textwidth}
\includegraphics[trim={0cm 0cm 0cm 0cm},clip, width= \textwidth]{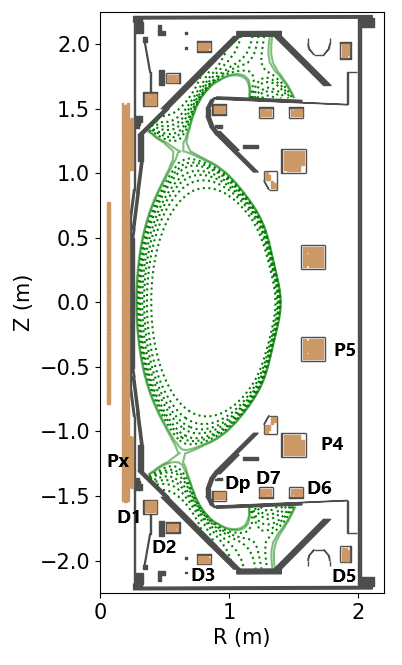}
\vspace{-0.8cm}
\caption{}
\end{subfigure}
\begin{subfigure}[b]{0.48\textwidth}
\includegraphics[trim={0cm 0cm 0cm 2cm},clip,width= \textwidth]{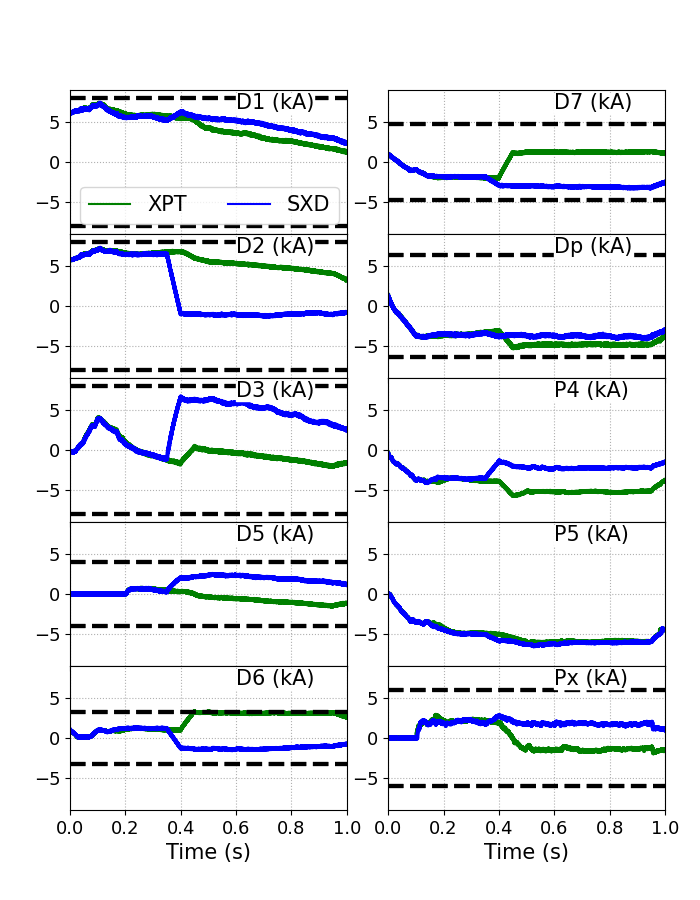}
\vspace{-0.8cm}
\caption{}
\end{subfigure}
\vspace{-0.4cm} 
\caption{(a) EFIT magnetic reconstruction of XPT discharge \#49320 at 0.5s with the separatrix plotted as a solid line and (b) comparison of poloidal field coil currents used to achieve the SXD and XPT configurations. The coil current limits are shown in black.}
\label{fig:fig01_currents}
\vspace{-0.1cm} 
\end{figure}

The XPT configuration studied in this work is obtained by taking an SXD reference discharge and introducing a 2nd poloidal field null in the divertor chamber, close to the separatrix. This results in a large poloidal flux expansion near the 2nd null, as can be seen from the magnetic reconstructions in figure \ref{fig:fig1_EFIT}.\\
The possible locations in which the 2nd null can be placed are limited by the coil currents achievable by the poloidal field coils, and the equilibrium studied in this work was designed to push the 2nd null position as close as possible to the center of the divertor chamber, while still being close enough to the separatrix to significantly affect the divertor state. The EFIT\cite{EFIT} magnetic reconstruction of the full equilibrium is plotted in figure \ref{fig:fig01_currents}. Also shown is a comparison of the poloidal field coils' currents used in the XPT and in the reference SXD discharge, highlighting how the XPT configuration can be obtained with coil currents of a comparable magnitude to those used for the SXD. \\
\begin{figure}[]\centering
\includegraphics[trim={0cm 0cm 0cm 2cm},clip,width= 0.75\textwidth]{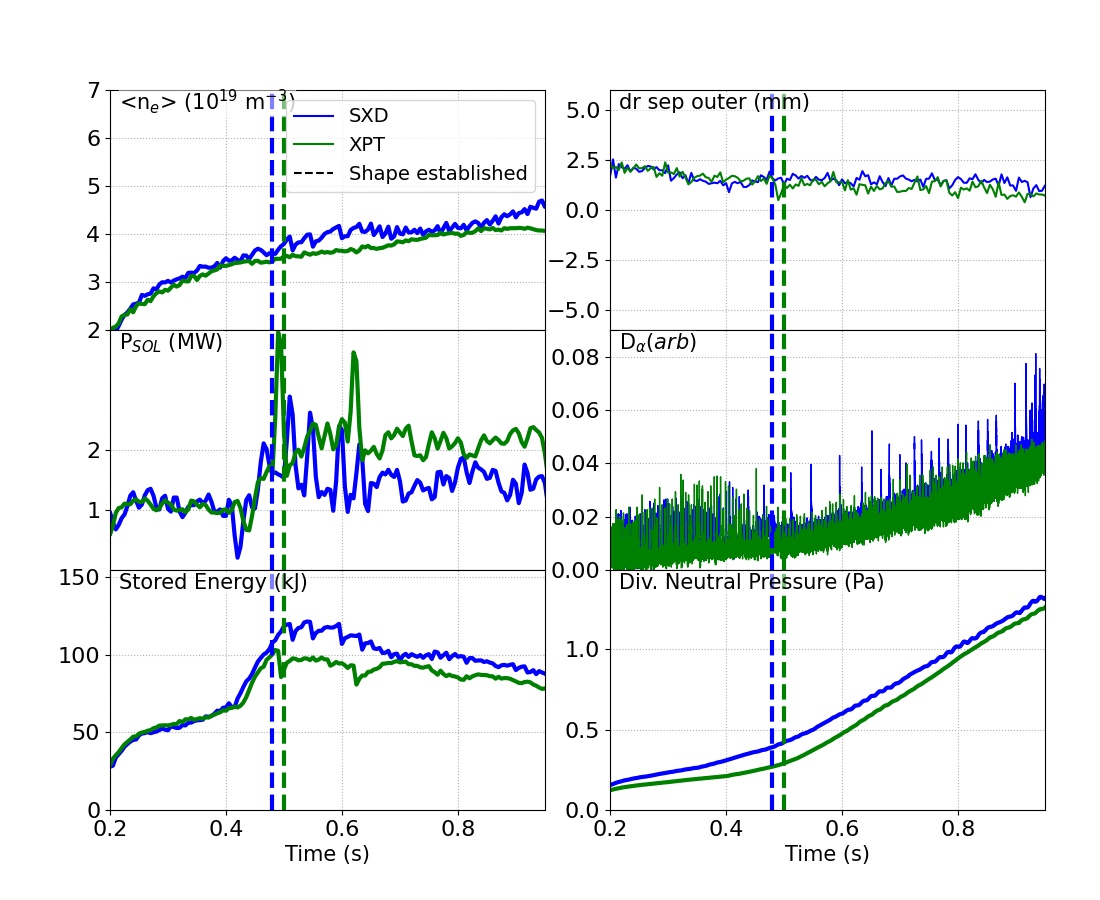}
\vspace{-0.5 cm}\caption[{Overview of SXD and XPT discharges, showing the line-averaged density, power crossing the separatrix ($P_{SOL}$), stored energy, distance between the primary and secondary separatrices (dr sep), midplane $D_\alpha$ signal, and divertor neutral pressure measured by the pressure gauge. }]{Overview of SXD and XPT discharges, showing the line-averaged density, power crossing the separatrix ($P_{SOL}$), stored energy, distance between the separatrices, midplane $D_\alpha$ signal, and divertor neutral pressure measured by the pressure gauge. }
\label{fig:fig8_overview}
\vspace{-0.3cm} 
\end{figure}
The reference SXD discharge is a type I ELMy H-mode and uses both NBI systems available on MAST-U with the maximum power currently available of $\sim 3.5$ MW. It also includes a divertor fuelling ramp to increasingly push the divertor into more deeply detached conditions and compare its performance as a function of divertor pressure.
The evolution of the XPT discharge (\#49320) is compared to the SXD discharge (\#49324) in figure \ref{fig:fig8_overview} in terms of some key discharge parameters. These include the line-averaged density, the power entering the scrape-off layer ($P_{SOL} = P_{Ohmic} + P_{NBI} - P_{\dd W/ \dd t} - P_{rad}$), the stored energy, the midplane $D_\alpha$ emission, the distance between primary and secondary separatrices (dr sep\footnote{Indicative of how magnetically balanced the double-null is, with a value of 0 indicating perfect balance and a value much smaller than the scrape-off layer width ($\sim$ 6 mm) being desired.}), and the divertor neutral pressure. The discharge ramp-up is performed in a conventional divertor configuration, and the shifts to the SXD and XPT configurations are completed at $\sim$ 0.5 s.  While the discharges start off in attached divertor conditions in the CD phase, they transition to the detached state as the strike point major radius is increased to establish the final divertor shapes. This is true even when operating with no divertor fuelling and no impurity seeding and is similar to previous observations in L-mode discharges in SXD configuration\cite{verhaegh2024improved}, where attached conditions could not be reached. 
Before establishing the divertor shape, the main plasma parameters are very similar between the two discharges. Some differences are seen after the shape is established, including a lower core density and stored energy in the XPT, as well as a higher $P_{SOL}$ and ELM frequency. Possible reasons and consequences of these differences are discussed further in section \ref{sec:XPT_caveats}.
\subsection{Evidence of 2nd null action in experimental discharges}
\begin{figure}[]\centering
\begin{subfigure}[b]{0.45\textwidth}
\scalebox{-1}[1]{\includegraphics[ width= \textwidth]{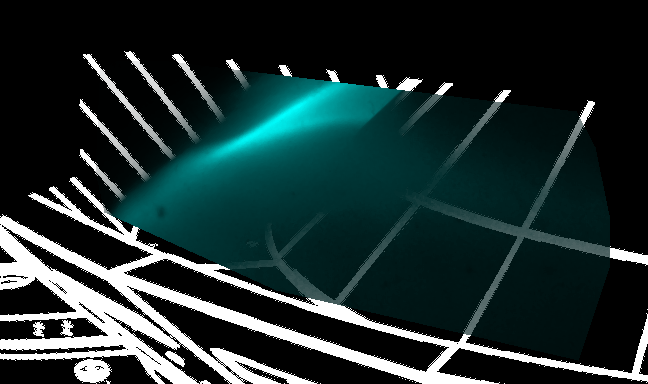}}
\caption{}
\end{subfigure}
\begin{subfigure}[b]{0.45\textwidth}
\scalebox{-1}[1]{\includegraphics[width= \textwidth]{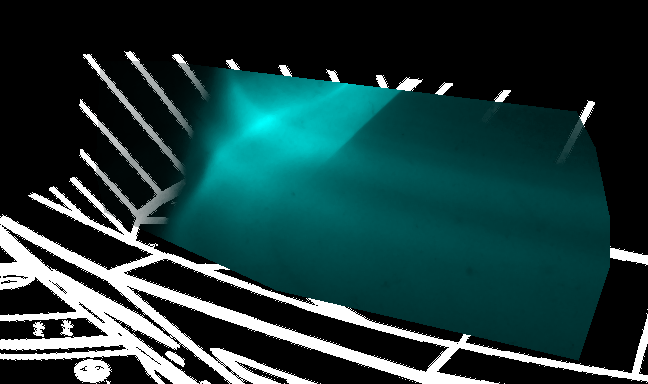}}
\caption{}
\end{subfigure}
\begin{subfigure}[b]{0.45\textwidth}
\includegraphics[ width= \textwidth]{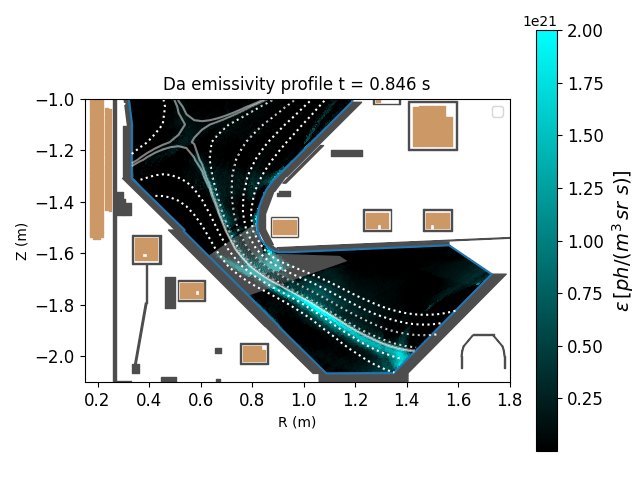} 
\vspace{-1cm}
\caption{}
\end{subfigure}
\begin{subfigure}[b]{0.45\textwidth}
\includegraphics[ width= \textwidth]{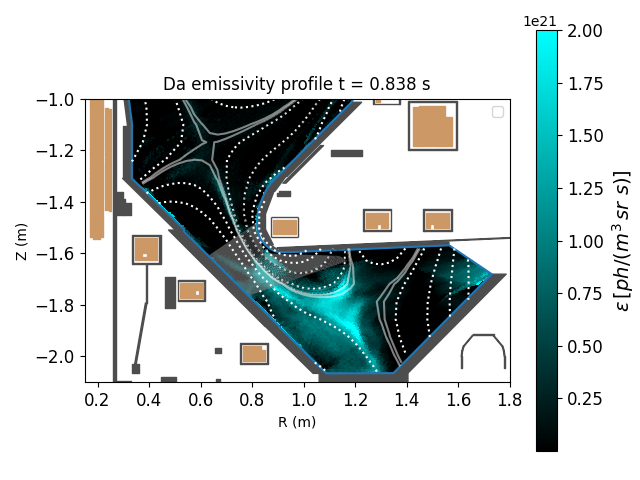} 
\vspace{-1cm}
\caption{}
\end{subfigure}
\vspace{-0.2cm} 
\caption{Balmer alpha inter-ELM camera images in the divertor chamber overlaid on a wireframe of the machine in (a) SXD and (b) XPT configuration. Corresponding 2D emissivity inversions in (c) SXD and (d) XPT configurations with flux surfaces overlaid in white. The emission near the baffle and the discontinuity in the emission near the divertor entrance are considered inversion artefacts due to the poor diagnostic coverage of the region. The region that can be affected by these artifacts is shaded in gray.}
\label{fig:fig5_MWI}
\vspace{-0.1cm} 
\end{figure}
In addition to the magnetic reconstruction, imaging data in the divertor can be used to verify that the achieved equilibrium has the 2nd null close enough to the separatrix to affect the divertor plasma. This is of particular importance in the XPT due to possible uncertainties in the magnetic reconstruction of the exact position of the 2nd null. Furthermore, the IR cameras cannot be used to measure the presence of significant heat flux to the secondary strike points due to the limited spatial coverage of the diagnostic, meaning that IR heat flux measurements are only available for the main strike point (e.g. for the lower divertor, only on the bottom of the divertor chamber and not on the underside of the baffle). Filtered $D_\alpha$ camera images of the divertor chamber from the MWI diagnostic\cite{feng_development_2021}\cite{wijkamp_MWI_2023} are shown in figure \ref{fig:fig5_MWI} in both configurations, where the presence of an "X" shape in the emission and the existence of a secondary leg pointing to the top of the divertor chamber are clearly visible in the XPT discharge. Tomographic inversions of the 2D emissivity profiles are also shown, with the magnetic reconstruction overlaid.
Two of the legs are visible in the tomographic reconstruction of the XPT configuration, one pointing towards the bottom of the divertor chamber and one towards the top, with comparable $D_\alpha$ emission. This suggests the 2nd null is significantly affecting the divertor plasma. In XPT, hydrogenic emission is usually seen on only three of the four legs\footnote{The separatrix joining the primary and secondary X-point is one of the legs, which then splits into three additional legs ending with three strike points in each outer divertor chamber.} in the divertor chamber, although some emission on the fourth leg can be seen during ELMs. It is worth noting how the emission is shifted towards the PFR ($\psi \approx 0.975$, visible also in figure \ref{fig:fig7_cross_field}) and not peaked at the separatrix in either configuration. As will be shown later in section \ref{sec:XPT_exp}, the peak heat flux is also peaked in the PFR, which possibly suggests magnetic reconstruction errors.\\
Having verified the successful development of an XPT configuration, its observed exhaust benefits are discussed next.
\section{Comparing the exhaust performance of the Super-X and X-point target divertor configuration}\label{sec:XPT_exp}
This section aims to describe the differences in the divertor plasma state in XPT  induced by the 2nd null and to study the resulting improved exhaust performance. These will be discussed in detail in the following sections and can be summarized as the XPT having:
\begin{itemize}
    \item Increased cross-field profile width, resulting in additional volume for plasma-neutral interactions.
    \item Decreased peak particle and heat fluxes at the target.
    \item Preliminary indications of improved ELM buffering performance.
\end{itemize}
\begin{figure}[h!]\centering
\begin{subfigure}[b]{0.45\textwidth}
\includegraphics[ width= \textwidth]{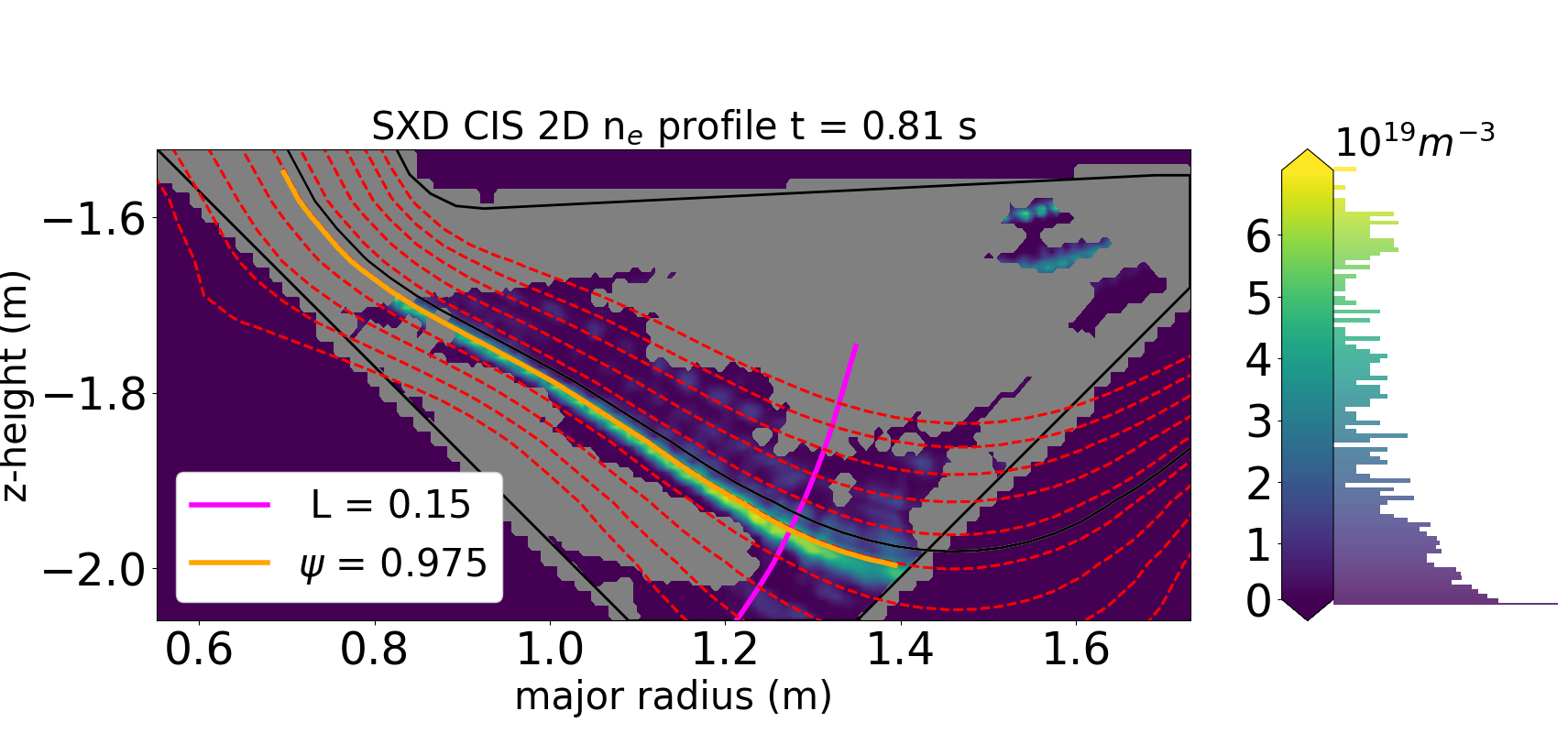} 
\caption{}
\end{subfigure}
\begin{subfigure}[b]{0.45\textwidth}
\includegraphics[width= \textwidth]{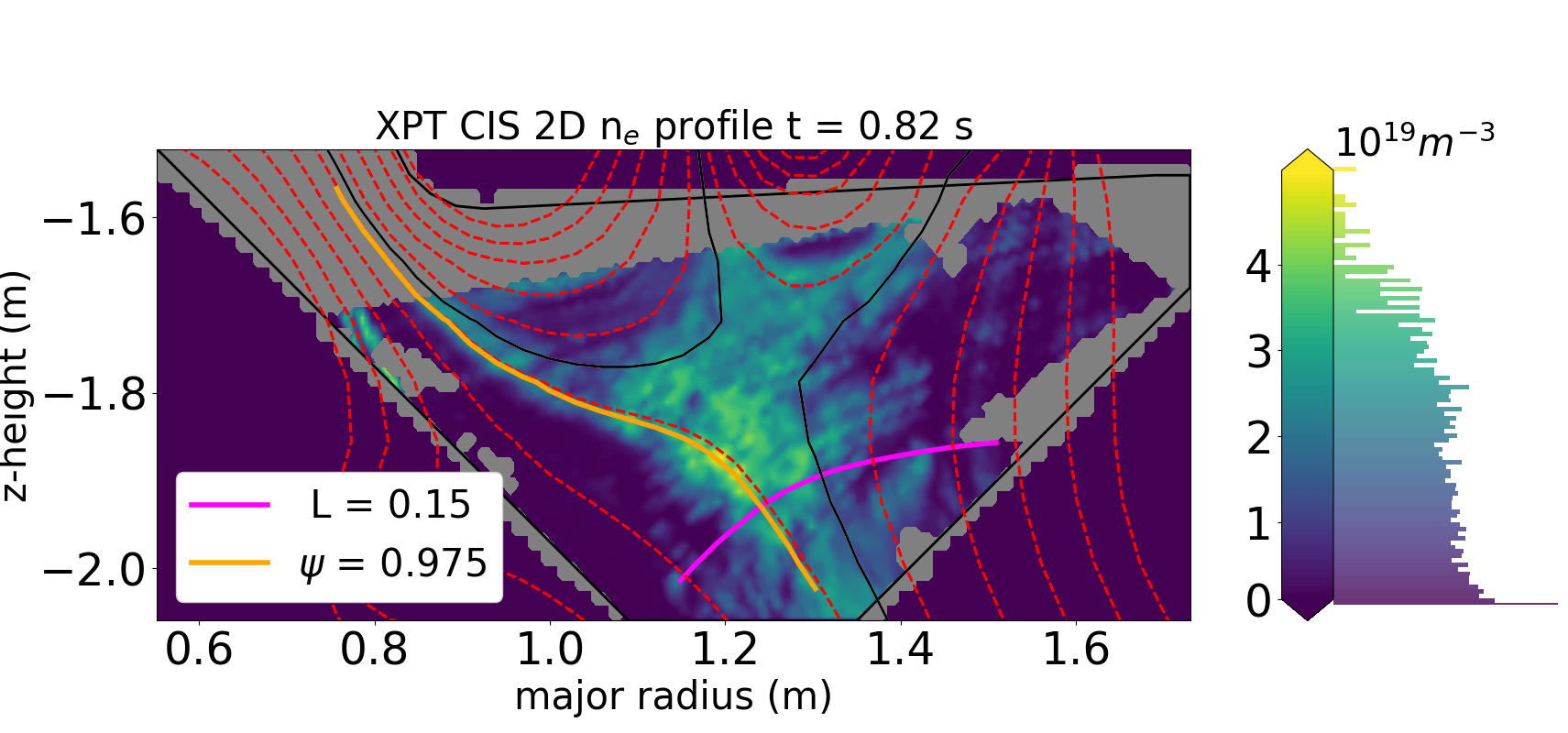}
\caption{}
\end{subfigure}
\vspace{-0.4cm} 
\caption{$n_e$ profiles inferred via coherence imaging spectroscopy in (a) SXD and (b) XPT configurations. Overlaid are the separatrix (black) and flux surfaces (red) from the EFIT magnetic reconstruction. The peak density is observed in the PFR and the $\Bar{\psi} = 0.975$ surface is shown in orange for reference. The cross-field line at 15 cm from the target (L = 0.15) along the separatrix is also shown, and the profiles along this line are compared in figure \ref{fig:fig7_cross_field}.}
\label{fig:fig6_CIS}
\vspace{-0.1cm} 
\end{figure}
\begin{figure}[h!]\centering
\begin{subfigure}[b]{0.45\textwidth}
\includegraphics[width= \textwidth]{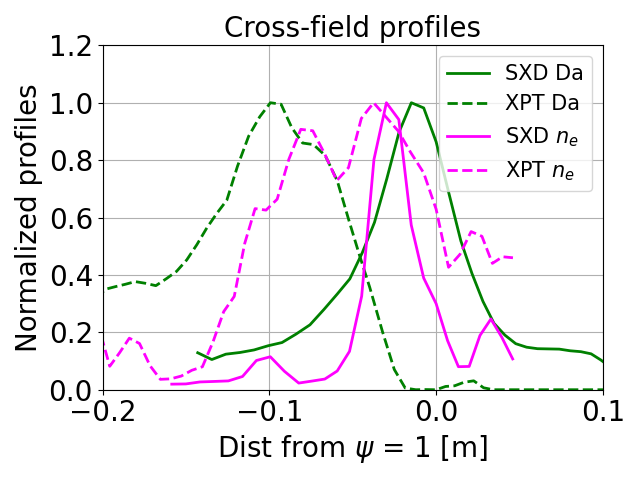}
\caption{}
\end{subfigure}
\begin{subfigure}[b]{0.45\textwidth}
\includegraphics[width= \textwidth]{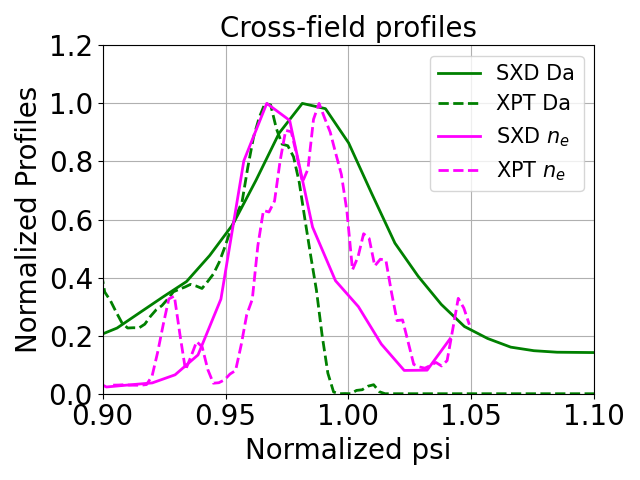}
\caption{}
\end{subfigure}
\vspace{-0.1 cm}\caption{(a) Comparison of normalized cross-field profiles of $n_e$ and $D_\alpha$ emission in the SXD and XPT discharges at 0.9 s and (b) the same profiles expressed mapped to $\psi$ coordinates. The cross-field profiles are taken at 15 cm from the target, along the magenta line shown in figure \ref{fig:fig6_CIS}.}
\label{fig:fig7_cross_field}
\vspace{-0.3cm} 
\end{figure}
\subsection{Broadening of the plasma-neutral interaction region}
The increased poloidal flux expansion near the 2nd null results in the plasma being spread out over a larger poloidal area. This is evidenced by the 2D divertor electron density profiles obtained via the coherence imaging diagnostic (CIS)\cite{Doyle_CIS}\cite{lonigro_CIS}, shown in figure \ref{fig:fig6_CIS}, where a broader $n_e$ profile is visible in the XPT case. Similarly to the $D_\alpha$ emission, the $n_e$ profile is peaked around $\psi \approx 0.975$, according to the EFIT reconstruction.\\
 This increase in width is compared directly using 1D cross-field profiles along trajectories that follow the gradient in $\psi$. The trajectories starting from a point with a 15 cm poloidal distance from the target along the separatrix are shown in magenta in figure \ref{fig:fig6_CIS} for both configurations, and the normalized profiles of $n_e$ and $D_\alpha$ emission along these lines are compared in figure \ref{fig:fig7_cross_field}.
While the width of both profiles is clearly larger in the XPT, the $n_e$ profile width becomes comparable when expressed as a function of normalized $\psi$. This indicates that if additional cross-field transport is present near the 2nd null, it does not result in significant additional broadening of the profile in addition to the expected broadening from the poloidal flux expansion. The $D_\alpha$ profile width appears wider in the SXD case when expressed as a function of $\psi$, which could be due to an underlying difference in neutral density distributions or possibly due to an artificial broadening of the SXD emission profile due to a limitation of the camera in reconstructing narrow emission profiles. This will be studied further in future work by comparing SXD configurations with increasing poloidal flux expansion. 
\begin{figure}[h!]\centering
\begin{subfigure}[b]{0.42\textwidth}
\includegraphics[trim={0cm 5.5cm 6.5cm 7cm},clip, width= \textwidth]{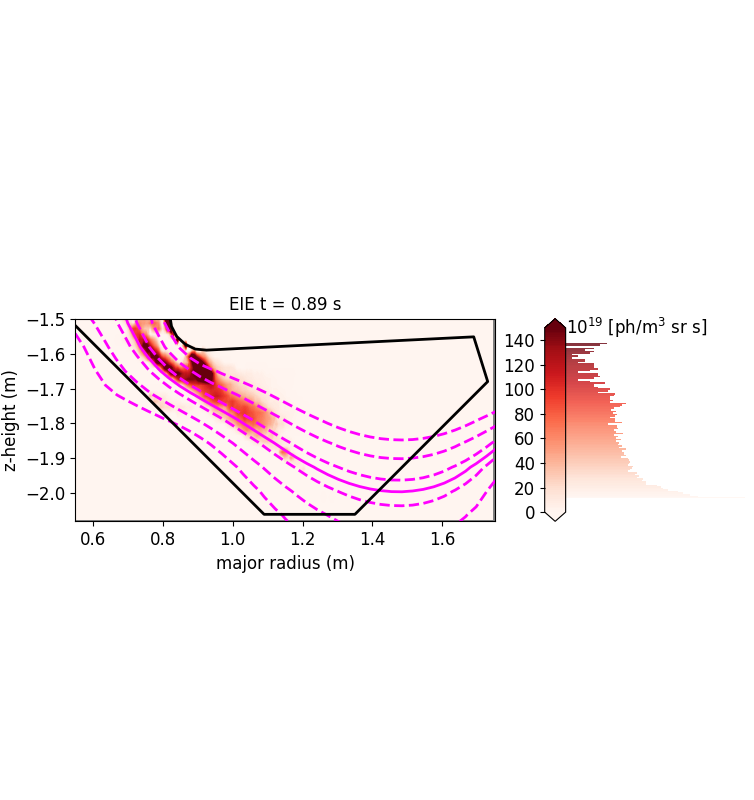} 
\caption{}
\end{subfigure}
\begin{subfigure}[b]{0.42\textwidth}
\includegraphics[trim={0cm 5.5cm 6.5cm 7cm},clip, width= \textwidth]{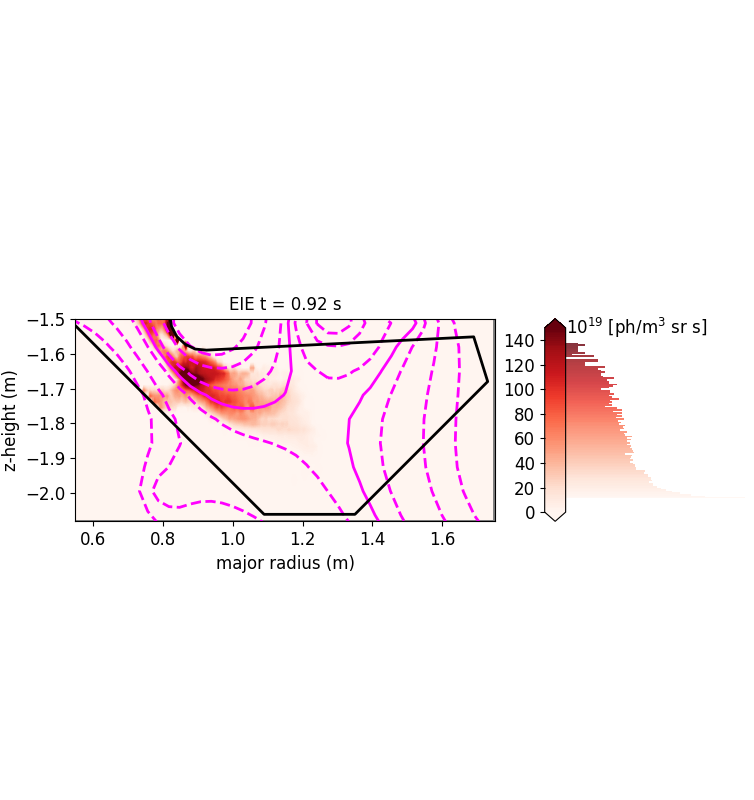} 
\caption{}
\end{subfigure}
\vspace{-0.75cm} 
\begin{subfigure}[b]{0.14\textwidth}
\includegraphics[trim={14.4cm 11.8cm 1.2cm 7cm},clip, width= \textwidth]{img/fig10a_EIE_SXD.png} 
\includegraphics[trim={12.7cm 5.5cm 4.76cm 8cm},clip, width= 0.5\textwidth]{img/fig10a_EIE_SXD.png} 
\end{subfigure}
\begin{subfigure}[b]{0.42\textwidth}
\includegraphics[trim={0cm 5.5cm 6.5cm 7cm},clip, width= \textwidth]{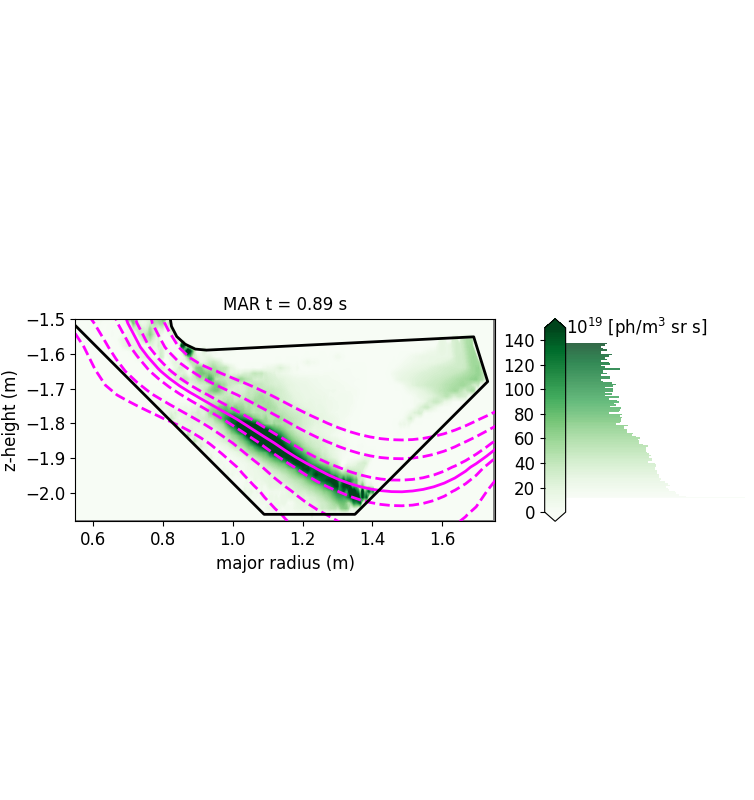} 
\caption{}
\end{subfigure}
\vspace{-0.75cm}
\begin{subfigure}[b]{0.42\textwidth}
\includegraphics[trim={0cm 5.5cm 6.5cm 7cm},clip, width= \textwidth]{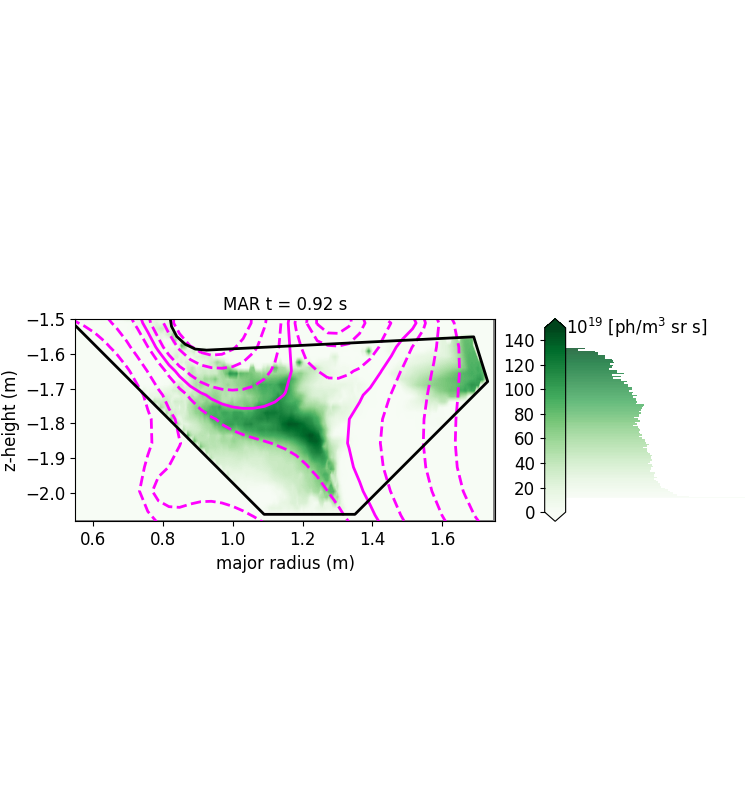} 
\caption{}
\end{subfigure}
\begin{subfigure}[b]{0.14\textwidth}
\includegraphics[trim={14.4cm 11.8cm 1.2cm 7cm},clip, width= \textwidth]{img/fig10b_MAR_SXD.png} 
\includegraphics[trim={12.7cm 5.5cm 4.76cm 8cm},clip, width= 0.5\textwidth]{img/fig10b_MAR_SXD.png} 
\end{subfigure}
\begin{subfigure}[b]{0.42\textwidth}
\includegraphics[trim={0cm 5.5cm 6.5cm 7cm},clip,width= \textwidth]{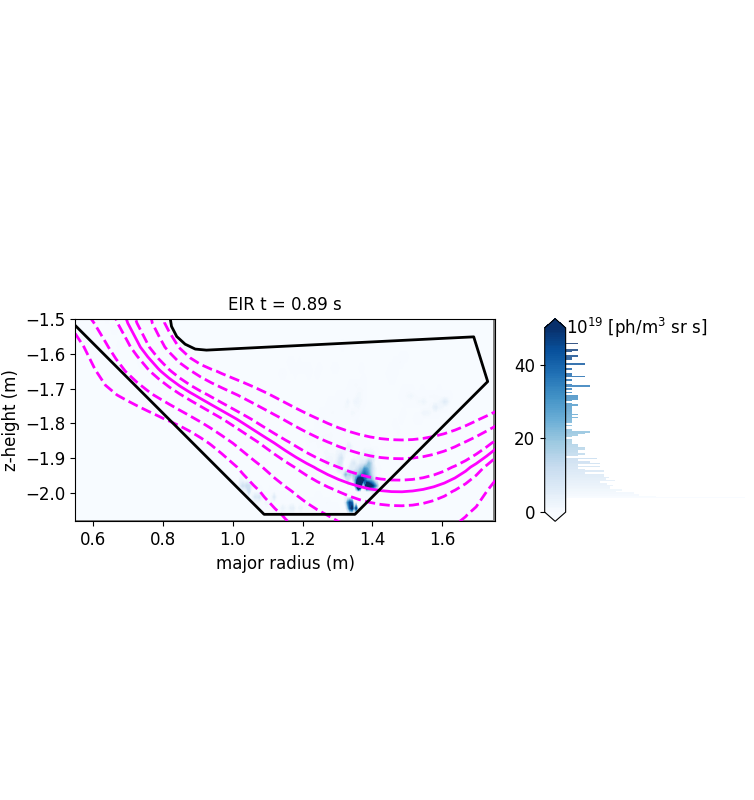}
\caption{}
\end{subfigure}
\begin{subfigure}[b]{0.42\textwidth}
\includegraphics[trim={0cm 5.5cm 6.5cm 7cm},clip,width= \textwidth]{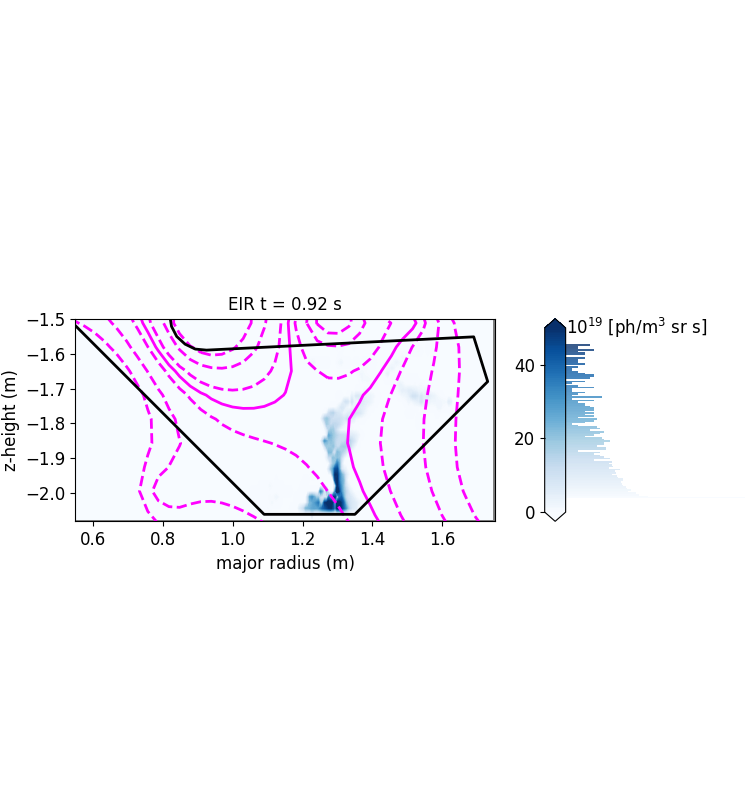}
\caption{}
\end{subfigure}
\begin{subfigure}[b]{0.14\textwidth}
\includegraphics[trim={14.4cm 11.8cm 1.2cm 7cm},clip, width= \textwidth]{img/fig10c_EIR_SXD.png} 
\includegraphics[trim={12.7cm 5.5cm 4.76cm 8cm},clip, width= 0.5\textwidth]{img/fig10c_EIR_SXD.png} 
\end{subfigure}
\vspace{-0.4cm} 
\caption{2D profile of $D_\alpha$ emission related to (a,b) EIE, (c,d) MAR and (e,f) EIR as inferred by the neural network in SXD (a,c,e) and XPT (b,d,f) configurations in the most deeply detached conditions reached ($P_0 \approx $ 1.2 Pa).}
\label{fig:fig10_NN_2D}
\vspace{-0.1cm} 
\end{figure}
\subsection{Volumetric processes}\label{sec:vol_processes}
\begin{figure}[h!]\centering
\includegraphics[ width= 0.66\textwidth]{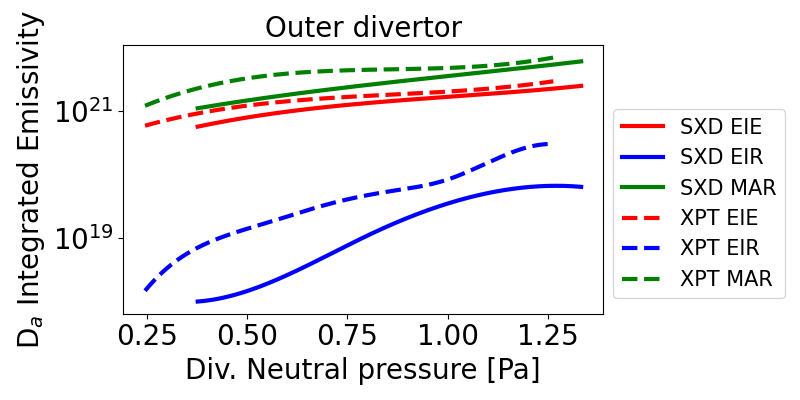} 
\vspace{-0.4cm} 
\caption{Volume integral in the lower outer divertor of the $D_\alpha$ emission due to EIE, MAR, and EIR emission in the SXD and XPT discharge. More recombination emission is inferred for the XPT and an earlier onset of EIR.}
\label{fig:fig11_em_int}
\vspace{-0.1cm} 
\end{figure}

The additional volume near and downstream of the 2nd null can lead to additional volumetric particle sinks and lower target temperatures. This is shown in figure \ref{fig:fig10_NN_2D}, where the 2D profiles of $D_\alpha$ emission attributed to electron-impact excitation (EIE), molecular activated recombination (MAR) and dissociation (MAD), and electron-ion recombination (EIR) by a neural network trained on the Fulcher, $D_\alpha$ and $D_\beta$ emission\cite{Lonigro_thesis} are shown for the two configurations for comparable divertor neutral pressure ($P_0 \approx 1.2$ Pa). Times near the end of the fuelling ramp are chosen to show the most detached conditions achieved.\\
The spatial distribution of the emission attributed to EIE, which is correlated to the ionization region \cite{Verhaegh_2023}, is comparable between the two discharges and upstream of the 2nd null. The divertor $D_\alpha$ emission is mostly attributed to plasma molecule chemistry (MAR and MAD), in particular in the proximity of the 2nd null in the XPT, which is also characterized by a wider emission profile, as shown in figure \ref{fig:fig7_cross_field}. Stronger EIR emission is inferred near the target in the XPT configuration, which is indicative of lower electron temperatures.\\
The volume integrals across the whole divertor for the different emission contributions are shown as a function of divertor neutral pressure in figure \ref{fig:fig11_em_int}.
Overall, the total emission in the divertor is larger in the XPT and is mostly attributed to the plasma-molecule interaction processes. Stronger $D_\alpha$ EIR emission is inferred in the XPT with respect to the SXD, particularly at lower divertor neutral pressures, although the EIR emission is still lower than the $D_\alpha$ emission from plasma-molecule interactions in the conditions reached in this discharge. This is partially attributed to the low electron densities reached with the currently available heating power ($<$ $6 \cdot 10^{19}$m$^{-3}$), and a more dominant role of EIR might be expected in higher power conditions as the electron density becomes higher \cite{lonigro20252delectrondensityprofile}. Stronger total radiation is also observed in the divertor chamber by the imaging bolometer\cite{MASTU_IRVB} in the XPT discharge ($\sim 150-200 \%$ of SXD value), although the diagnostic integrates in time over the ELMs and the difference could also be due to the differences in core scenario (section \ref{sec:XPT_caveats}).
\subsection{Target quantities}\label{sec:XPT_target}
\begin{figure}[h!]\centering
\includegraphics[ width= 0.75\textwidth]{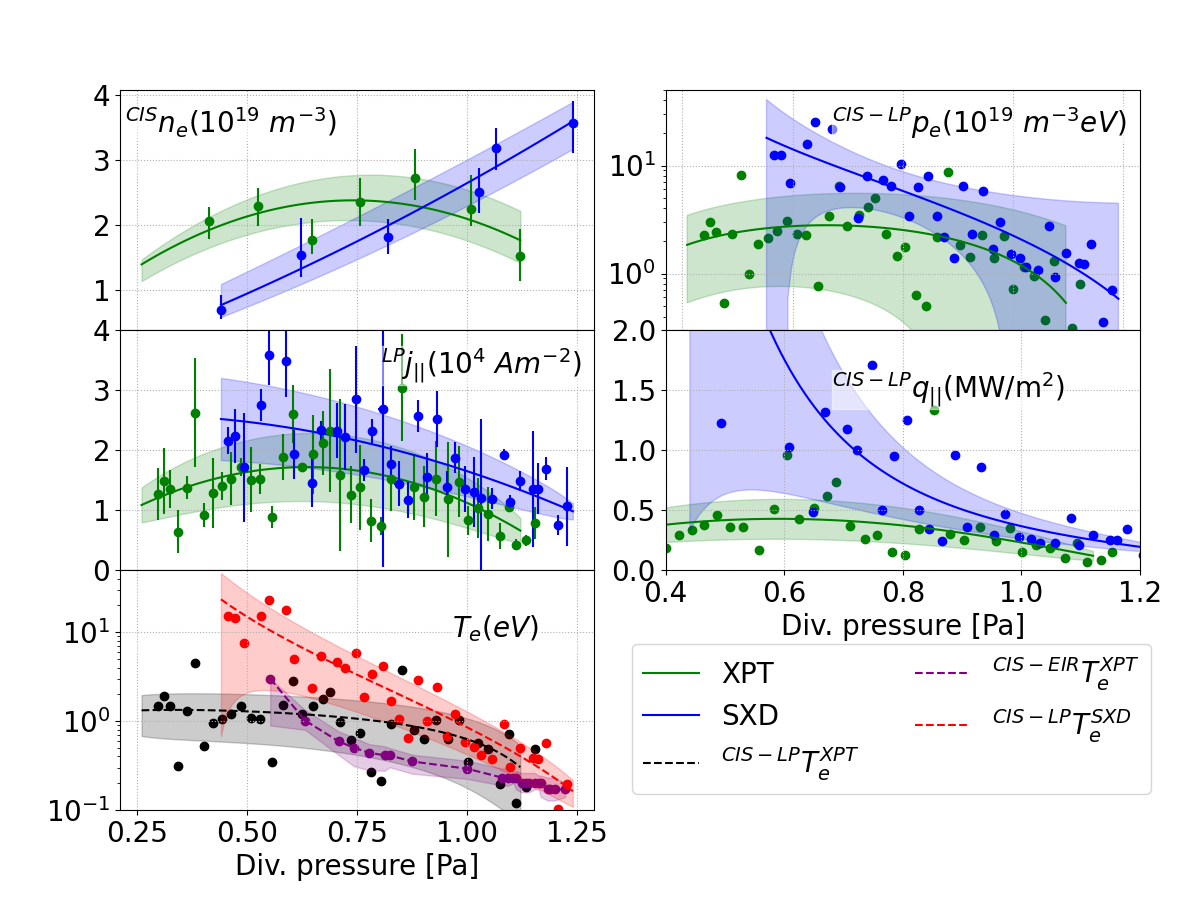} 
\vspace{-0.4cm} 
\caption{Target $n_e$ inferred by CIS along the $\psi = 0.975$ flux surface, peak particle flux inferred by the Langmuir probes, and $^{CIS-LP}T_e $ and $^{CIS-EIR}T_e$ estimates in the (a) SXD and (b) XPT discharges. The green points are obtained by combining the LP data with a fit of the CIS $n_e$  data, while the yellow points by combining the neural network EIR estimates with a fit of the CIS $n_e$ data to show the extent of the scatter in the underlying data. The uncertainty bounds of the $n_e$ and $j_{||}$ traces are obtained as the standard deviation of the measured points with respect to the polynomial fit, while the uncertainties on the $T_e$ inferences are propagated from that. Peak target heat flux deposited by the plasma inferred by combining the particle flux measured by the Langmuir probes and the $T_e$ estimates in SXD and XPT discharges.}
\label{fig:fig12_Te}
\vspace{-0.1cm} 
\end{figure}

The additional power, momentum, and particle losses expected in the XPT configuration should lead to a lower $T_e$, peak particle flux, and peak heat flux reaching the divertor target. If these quantities were peaked at the separatrix inferred by the EFIT magnetic reconstruction, they would be peaked on the underside of the baffle, on the top of the divertor chamber, which is outside the coverage of the divertor diagnostics. As discussed further in section \ref{sec:XPT_caveats}, however, the peak heat flux in both SXD and XPT configurations is peaked at $\psi \approx 0.975$, in agreement with the 2D CIS $n_e$ measurements in figure \ref{fig:fig6_CIS} and the $D_\alpha$ emission measurements in figure \ref{fig:fig5_MWI} in the lower divertor, and thus in view of the divertor diagnostics. This shift from $\psi = 1$ points to an error in the magnetic reconstruction, which will be investigated further in future work.\\
The peak ion saturation current at the target measured by Langmuir probes (LP) and the electron density inferred by CIS at 3 cm from the target\footnote{The electron density from the 2D profile is not taken exactly at the target to avoid inversions artifact that can develop in the last few centimeters closest to the target \cite{lonigro_CIS}} along the $\psi=0.975$ flux surface are compared in figure \ref{fig:fig12_Te} for the two configurations as a function of divertor neutral pressure. The divertor Thomson scattering could not be used to compare the target $T_e$ due to the poor alignment with the strike point, particularly in XPT. Furthermore, target $T_e$ measurements from Langmuir probes are not reliable in these strongly detached conditions, but current density measurements ($j_{||}$) are. Thus, $T_e$ inferences are obtained using two alternative methods: either by combining the CIS and LP measurements ($^{CIS-LP}T_e $) or by combining the $D_\alpha$ EIR emission and the CIS measurements ($^{CIS-EIR}T_e$) by solving respectively equations \ref{eq:CIS_LP} and \ref{eq:CIS_EIR}\cite{lonigro20252delectrondensityprofile}. 
\begin{align}
    & ^{CIS-LP}T_e = \frac{\exp(1)m_i}{2e}\frac{^{LP}j_\perp^2}{^{CIS}n_e^2} \label{eq:CIS_LP}\\
    & \epsilon_{D_\alpha,EIR}(^{CIS}n_e, ^{CIS-EIR}T_e) = ^{CIS}n_e^2PEC_{rec}(^{CIS}n_e, ^{CIS-EIR}T_e) \label{eq:CIS_EIR}
\end{align}
with the $PEC_{rec}$ effective coefficients for EIR emission taken from the ADAS\cite{ADAS} database.\\
Both these estimates are also shown in figure \ref{fig:fig12_Te}, although the $^{CIS-EIR}T_e$ is only shown for the later stages of the XPT discharge, as it loses sensitivity at high temperatures due to the negligible EIR emission, and similarly no $^{CIS-EIR}T_e$ estimates were possible in the SXD due to the negligible EIR emission on the $\psi = 0.975$ flux surface.
Both the peak particle flux and the target $T_e$ are smaller in the XPT configuration, although the difference becomes less significant at higher divertor neutral pressures as the temperatures converge to $\sim$ 0.3 eV in both cases. Reasonable agreement is found between the two $T_e$ estimates. The $n_e$ profile is observed to continuously rise near the target in the SXD case, even though the divertor is already detached at the lowest divertor pressures. This is consistent with previous L-mode results \cite{lonigro20252delectrondensityprofile} and attributed to momentum losses in the detached region due to plasma-molecule collisions slowing down the plasma flow. The XPT case does not show a similar increase in $n_e$ with deepening detachment, which could be expected given the stronger ion sink contributions, but further measurements at higher heating powers (and thus, higher maximum divertor $n_e$) would be needed to study this further. \\
From the $n_e$ and $T_e$ estimates, the electron pressure at the target $p_e = n_e T_e$ can be estimated. The decreasing pressure with increasing neutral pressure is evidence of increasing momentum losses, while the lower electron pressure in the XPT for similar divertor neutral pressure is evidence of additional momentum losses in the XPT compared to the SXD, particularly for lower divertor neutral pressures.
Similarly, from the LP particle fluxes and a $T_e$ estimate, the peak parallel heat flux carried by the plasma when reaching the target, accounting for surface recombination, is given by 
\begin{align}
    q_{||} = (\gamma T_e + E_{rec})\Gamma_{||}
\end{align}
with $\gamma\sim 7.5$ the sheath transmission factor\cite{Stangeby_2018}  and $E_{rec} = 13.6 + 2.2$ eV the energy deposited via surface recombination, accounting for both the atomic and molecular potential energy. For $T_e < 2$ eV, most of the heat flux is deposited via surface recombination and the analysis results are only weakly sensitive to the $T_e$ estimates, especially for $T_e < 1$ eV. These ``plasma" heat fluxes inferred using the $^{CIS-LP}T_e$ estimates are also compared for the two configurations in figure \ref{fig:fig12_Te}. 
Significantly smaller heat fluxes are inferred in the XPT configuration, particularly at low divertor pressures. The measurements appear to converge at higher divertor pressures, but this may be due to a lower bound on the measurement capabilities of the diagnostics, given the very low heat flux values. The $q_\perp$ reduction (not shown) is even stronger, due to the smaller angle between the flux surfaces and the target in XPT.\\
While the IR heat flux measurements also show lower values in XPT (not shown), significant discrepancies with the estimates in figure \ref{fig:fig12_Te} are found in deeply detached conditions, with $q_\perp$ values measured by the IR cameras more than 4 times higher than the plasma heat flux estimates in the later stages of the SXD discharge. These could be attributed to the reduced reliability of IR measurements in strongly detached conditions, attributed to the contamination of the IR signal by volumetric radiation, or to the total heat flux to the target being dominated by the neutrals and the radiation, as was also reported for SOLPS simulations of the MAST-U SXD in strongly detached Ohmic conditions\cite{moulton_super-x_2024}.
\subsection{ELM buffering}\label{sec:XPT_ELMs}
\begin{figure}[h!]\centering
\begin{subfigure}[b]{0.49\textwidth}
\includegraphics[ width= \textwidth]{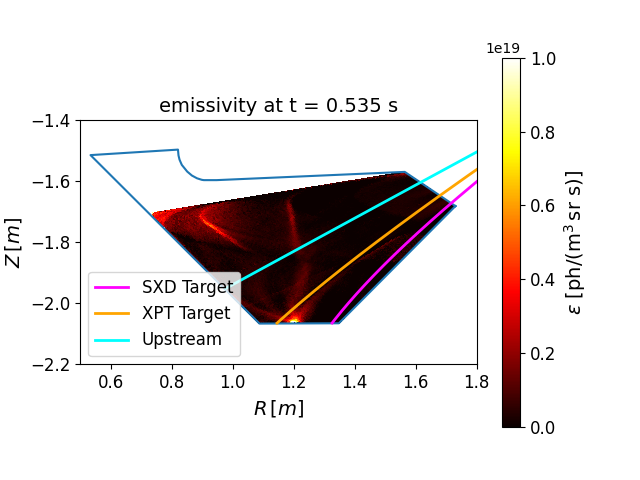} 
\caption{}
\end{subfigure}
\begin{subfigure}[b]{0.49\textwidth}
\includegraphics[width= \textwidth]{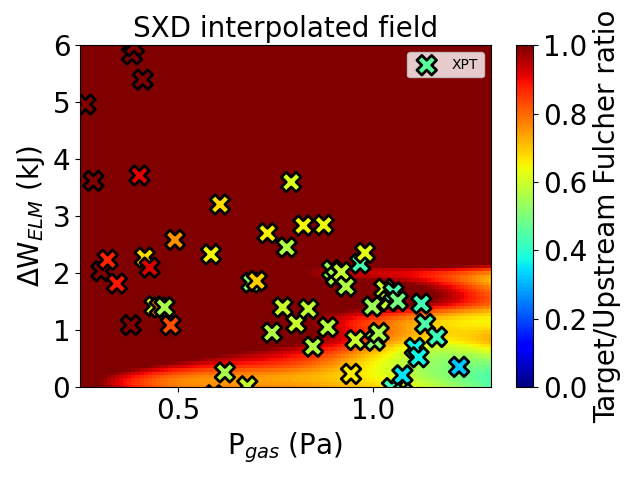}
\caption{}
\end{subfigure}
\vspace{-0.4cm} 
\caption{(a) Spectroscopy lines of sight used for the Target/Upstream burn-through proxy overlaid on 2D Fulcher emissivity profile in XPT configuration, integrating over an ELM which burns through to the target at low divertor pressure. Different lines of sight are used for the target in SXD and XPT configurations due to the different strike point locations. (b) Upstream/Target burn through proxy over the 2D parameter space of divertor neutral pressure and ELM energy. The 2D background color is obtained by interpolating the SXD data over the 2D space. The presence of lightly colored crosses over a darker background is indicative of better ELM buffering performance in the XPT configuration.}
\label{fig:fig15_burnthough}
\vspace{-0.1cm} 
\end{figure}

The additional volume available for plasma-neutral interactions in the XPT topology could also be useful in improving the divertor's ability to buffer transients, such as ELMs, which can burn through the detached region and re-attach the divertor momentarily. The SXD has already shown the ability to buffer small ELMs, in conditions in which the conventional divertor is attached even before the ELM, and for this buffering capability to increase with divertor neutral pressure\cite{Flanagan_2025}. A similar analysis can be performed in the XPT case to check if the parameter space in which the ELMs are buffered is larger in this configuration. \\
To determine if the ELM is buffered, a simple proxy consisting of the ratio between the Fulcher emission near the target and further upstream can be used. Using the Fulcher emission as a proxy for the ionization region, strong emission near the target is indicative of divertor re-attachment during the transient. To measure the Fulcher emission with a high time resolution, filtered photodiodes (part of the Ultra Fast Divertor Spectroscopy\cite{Flanagan_2025} diagnostic) integrating along a line of sight close to the target and a line of sight further up the divertor chamber ("upstream") can be used, with a ratio $>1$ indicative of reattachment. A more complete description of this simple proxy for ELM buffering can be found in the original work in the SXD divertor\cite{Flanagan_2025}. The lines of sight used for this metric are overlaid on a 2D Fulcher emissivity profile, which itself has an integration time that is too long to resolve the burnthrough process and thus results in an emissivity-weighted time-average of the steady-state and ELM emission, in figure \ref{fig:fig15_burnthough}. Different spectroscopy lines of sight are used for the target due to the different strike point positions in the two configurations. The experimental ratio for ELM data gathered during the XPT discharge and comparable SXD discharges is plotted in figure \ref{fig:fig15_burnthough}b as a function of ELM energy ($\Delta W_{ELM}$, calculated as the drop in stored energy based on on EFIT estimates with a time resolution of $\sim 0.6$ ms.) and divertor neutral pressure measured by the gauge. The background color is obtained as a 2D interpolation of the target/upstream ratio in 9 SXD discharges with a mix of ELM energies, while the crosses show XPT data, again colored according to the target/upstream ratio in XPT. Improved buffering of the XPT cases is demonstrated by the crosses being colored more towards the lower end of the color bar spectrum compared to the background on which they sit.
Both configurations see the ELMs burning through at low divertor pressure ($\leq$ 0.4 Pa), and both buffer the ELMs (ratio $<1$) for high neutral pressure ($>1$ Pa) and low ELM energy. However, the XPT also shows buffering for intermediate divertor pressures where the ELMs burn through in SXD, suggesting that the buffering is improved in the XPT.
These results should be considered qualitative, as quantitative studies using the UFDS in XPT configuration are challenging due to the 2D nature of the geometry and the effects of line-integration along the line of sight of the photodiode, which could affect the interpretation of the UFDS data. Furthermore, both the magnetic geometry and the position of the peak heat flux may change during the ELM. Future work will rely on fast cameras currently being installed on MAST-U to characterize the behaviour of the ELMs in 2D and with improved coverage of the leg going to the underside of the baffle.

\section{Discussion}\label{sec:discussion}
Having discussed some of the observed improvements in the XPT configuration, comparisons with reduced models are used to explore potential physics theories behind these benefits and how the XPT behaviour would extrapolate to more attached conditions.
Some limits of the current experimental scenario are then highlighted, before discussing the implications of these results.
\subsection{The effect of increasing connection length}\label{sec:XPT_parallel_profile}
\begin{figure}[]\centering
\begin{subfigure}[b]{0.45\textwidth}
\includegraphics[  width= \textwidth]{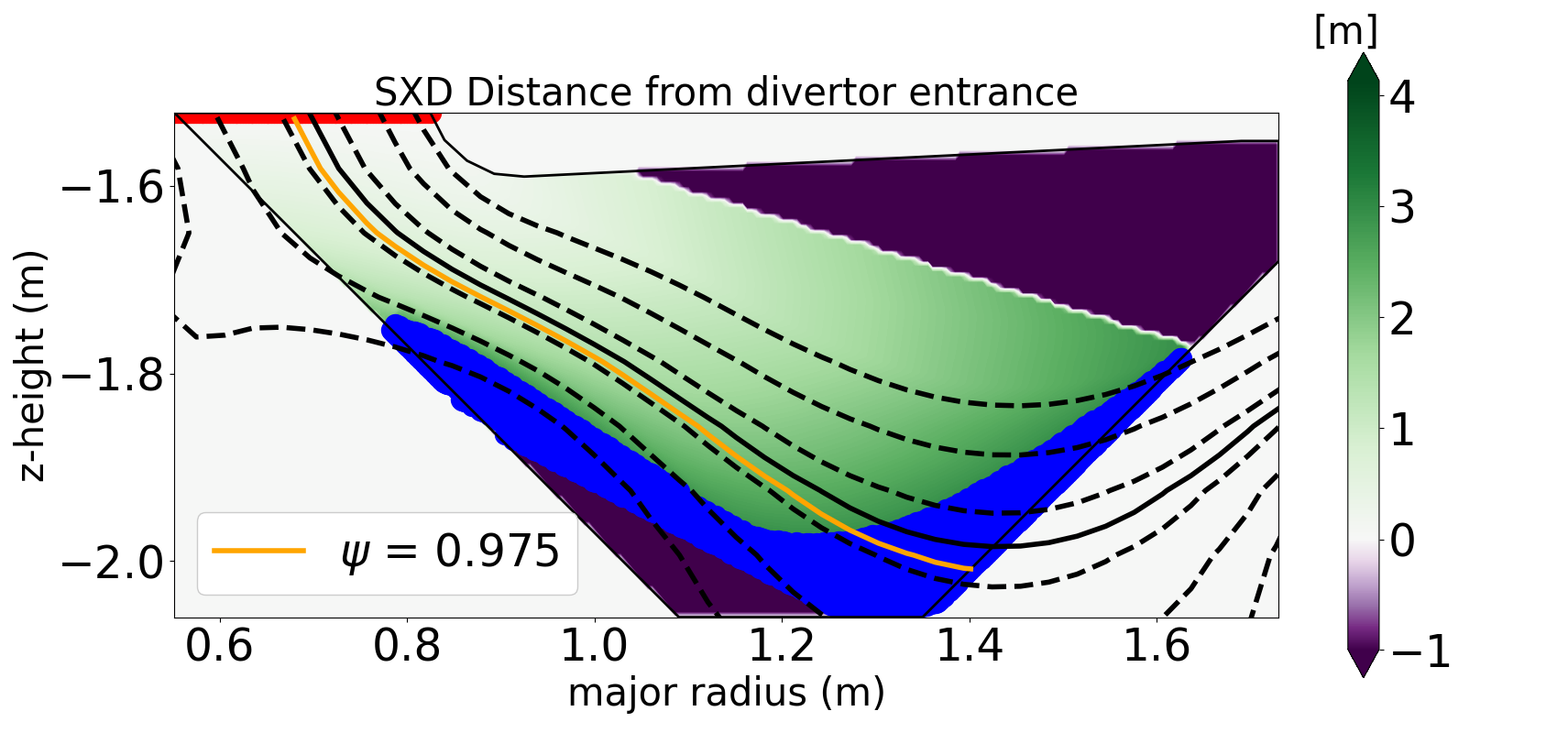} 
\caption{}
\end{subfigure}
\begin{subfigure}[b]{0.45\textwidth}
\includegraphics[ width= \textwidth]{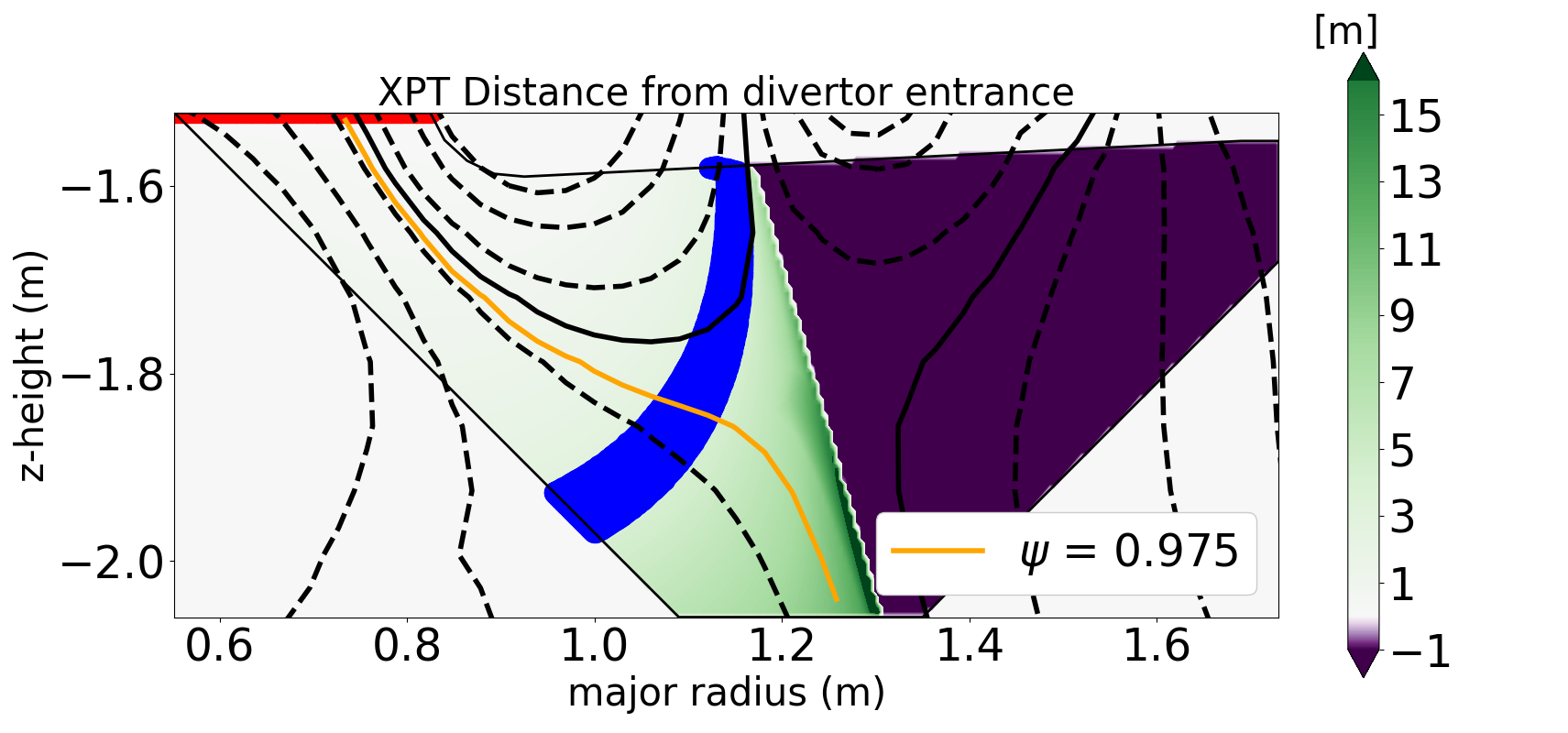}
\caption{}
\end{subfigure}
\caption{2D profiles of parallel distance from the divertor chamber entrance in (a) SXD and (b) XPT configuration. The area shaded in blue shows the region with a distance between 3 and 4 meters from the divertor entrance, highlighting how the 2nd null in XPT is at a parallel distance comparable to the target in SXD. The flux surface a $\psi = 0.975$ is also shown for reference, as the flux surface with the peak heat flux in both configurations. Areas not reached by the tracing are set to negative values. The starting points for the tracing are shown in red. The color bar is capped in the XPT case, where the maximum value of the tracing is limited by the spatial resolution of the magnetic reconstruction.}
\label{fig:fig2_dist}
\end{figure}

\begin{figure}[h!]\centering
\includegraphics[ width= 0.5\textwidth]{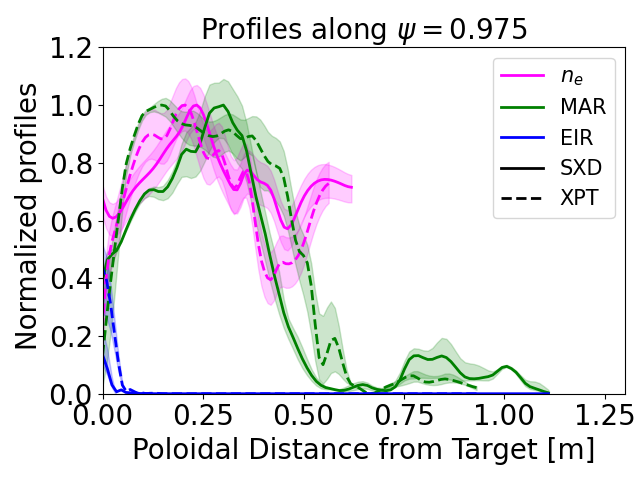} 
\vspace{-0.4cm} 
\caption{Normalized poloidal profiles along the separatrix of $n_e$ and the $D_\alpha$ recombination emission contributions in SXD and XPT configurations, highlighting the comparable spatial profiles when expressed in terms of poloidal distance from the target. The $n_e$ profiles are normalized to their maximum, while the emission profiles are normalized to the maximum $D_a$ emission along the profile. A higher EIR emission contribution can be observed in the XPT, as well as a wider MAR profile along the field.}
\label{fig:fig014_NN_along_sep}
\vspace{-0.1cm} 
\end{figure}

 One of the effects of introducing the 2nd null in the divertor chamber is the further increase in parallel distance to the target due to the null poloidal field near the X-point. This is shown in figure \ref{fig:fig2_dist}, in which the parallel distance to the divertor chamber entrance is compared for the SXD and XPT. 
The parallel distance in XPT is significantly larger, and the same parallel distance to the target in SXD corresponds roughly to the position of the 2nd null in XPT. This should lead to lower target electron temperatures in conduction-dominated regimes for the XPT, according to reduced models such as the two-point model\cite{Stangeby_2018}. Additional dissipation in the detached regime could also be expected through the additional plasma-neutral interactions that can happen in the detached region of the divertor leg, now larger in parallel space,  as the number of collisions undergone by the plasma before reaching the target is increased. \\
The two configurations thus have comparable poloidal leg length, but different connection length. A comparison of the poloidal profiles can then be instructive to distinguish the effects of parallel and poloidal leg length on the plasma profiles. The normalized profiles along the $\psi \sim 0.975$ flux surface of $n_e$, the MAR $D_\alpha$ emission, and the EIR $D_\alpha$ emission are compared for the two configurations in figure \ref{fig:fig014_NN_along_sep} for high divertor neutral pressures ($P_0 \approx $ 1.2 Pa, also shown in figure \ref{fig:fig10_NN_2D}). 
The normalized profiles look remarkably similar when expressed in terms of poloidal distance, although the XPT discharge has stronger EIR-related emission at the target. The MAR-related emission is less peaked in the XPT case, showing a broader emission profile along the field (as well as across it, as previously shown for the total $D_\alpha$ emission in figure \ref{fig:fig7_cross_field}).\\
These results seem to suggest that the poloidal leg length is the main driver of the plasma profiles in the detached region; if, instead, the additional parallel distance downstream of the secondary null resulted in additional power and momentum losses via plasma-neutral collisions, an upstream shift of the profiles in the XPT discharge might be expected. While this is somewhat observed in the recombination emission profiles, it is not present in the $n_e$ profile. This could suggest a different effect on the parallel $T_e$ and $n_e$ profiles, an influence of the poloidal distance to the walls, or might also be due to the differences in the core behavior discussed in section \ref{sec:XPT_caveats} and would require further discharges to test its reproducibility, for example also comparing different SXD geometries with varying poloidal flux expansion.
\subsection{Comparison of detachment location sensitivity}\label{sec:XPT_DLS}
\begin{figure}[h!]\centering
\begin{subfigure}[b]{0.52\textwidth}
\includegraphics[ width= \textwidth]{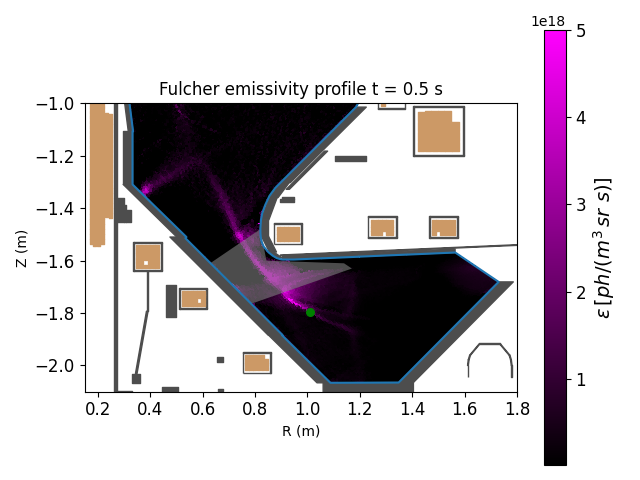}.  
\caption{}
\end{subfigure}
\begin{subfigure}[b]{0.45\textwidth}
\includegraphics[width= \textwidth]{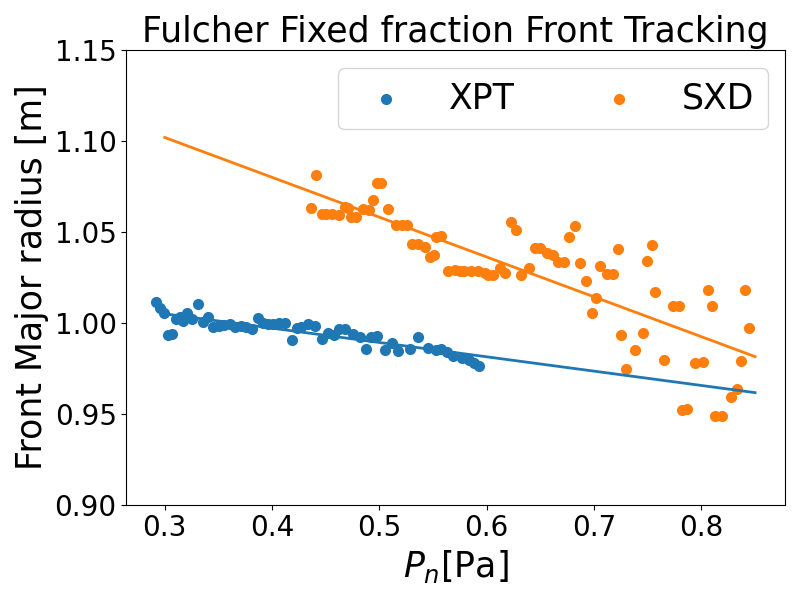}
\caption{}
\end{subfigure}
\vspace{-0.4cm} 
\caption{(a) 2D Fulcher emissivity profile with no divertor fuelling in XPT discharge \#49320. The area of the 2D profile affected by an inversion artifact due to a lack of diagnostic coverage is overlaid in gray. The front position is highlighted as a green dot. (b) Comparison of the Fulcher front position during the divertor fuelling ramp, highlighting the similar slope of movement between the SXD and XPT configurations upstream of the 2nd null. A linear fit is also shown as a visual aid.}
\label{fig:fig9_fonts}
\vspace{-0.1cm} 
\end{figure}
Consistent with previous SXD L-mode\cite{verhaegh2024improved} and H-mode\cite{Harrison_2024} results, both SXD and XPT divertor configurations are strongly detached in these discharges, even using the highest NBI power currently available on the machine and without any extrinsic impurity seeding or divertor fuelling.  Measurements of the $D_2$ Fulcher emission front (50 \% of the maximum along the separatrix) are routinely used on MAST-U as a proxy for the ionization (and thus detachment) front position \cite{verhaegh_molecules_2_2021}\cite{kool2024demonstrationsuperxdivertorexhaust}\cite{moulton_super-x_2024}. The 2D emissivity reconstruction of the $D_2$ Fulcher band emission in the divertor is shown in figure \ref{fig:fig9_fonts}a at the start of the $D_2$ divertor fueling ramp in XPT configuration, showing the most attached conditions achieved in the discharge. The Fulcher emission front has a poloidal distance from the target of $\sim$0.4 m along the reference fieldline ($\psi$ = 0.975), indicative of detached conditions, and is upstream of the 2nd null in the XPT discharge. As such, a comparison of the detachment front sensitivity downstream of the 2nd null is not possible. The location of the Fulcher emission front in both discharges is shown in figure \ref{fig:fig9_fonts}b as a function of divertor pressure\footnote{The front position is plotted here as a function of the major radius to avoid shifts due to the different X-point positions in the two discharges when plotting as a function of poloidal distance from the X-point. Due to the larger X-point radius in the XPT discharge, it would result in the XPT having a front further upstream (closer to the primary X-point) compared to the SXD, although the spatial location of the emission front in the divertor is comparable.}.
 The detachment front in XPT is shifted slightly upstream of the SXD one, which might suggest additional dissipation in the XPT, but might also be due to differences in the core scenario, discussed more in section \ref{sec:XPT_caveats}. The location sensitivity (i.e. slope) is comparable between the two configurations and possibly even reduced in the XPT in the narrow movement range achieved in this discharge. The comparable slope in the front movement suggests that by combining the total-flux expansion of the SXD and the second poloidal field null of the XPT,  the resulting geometry still retains all the benefits of the SXD regarding the reduced sensitivity of the detachment front location when the detachment front is upstream of the 2nd null and is thus unaffected by it. This is an example of how multiple ADC strategies can be combined to further increase their benefits. \\
On TCV XPT discharges, the detachment front location sensitivity is strongly reduced in the proximity of the 2nd null\cite{Lee_XPT}. While this could not be observed yet on MAST-U due to the strongly detached conditions, the experimental MAST-U equilibria can be used to make qualitative predictions on the expected behaviour using reduced models.
\subsubsection{Extrapolations with the detachment location sensitivity model}\label{sec:XPT_DLS_theory}
\begin{figure}[]\centering
\includegraphics[width= 0.7\textwidth]{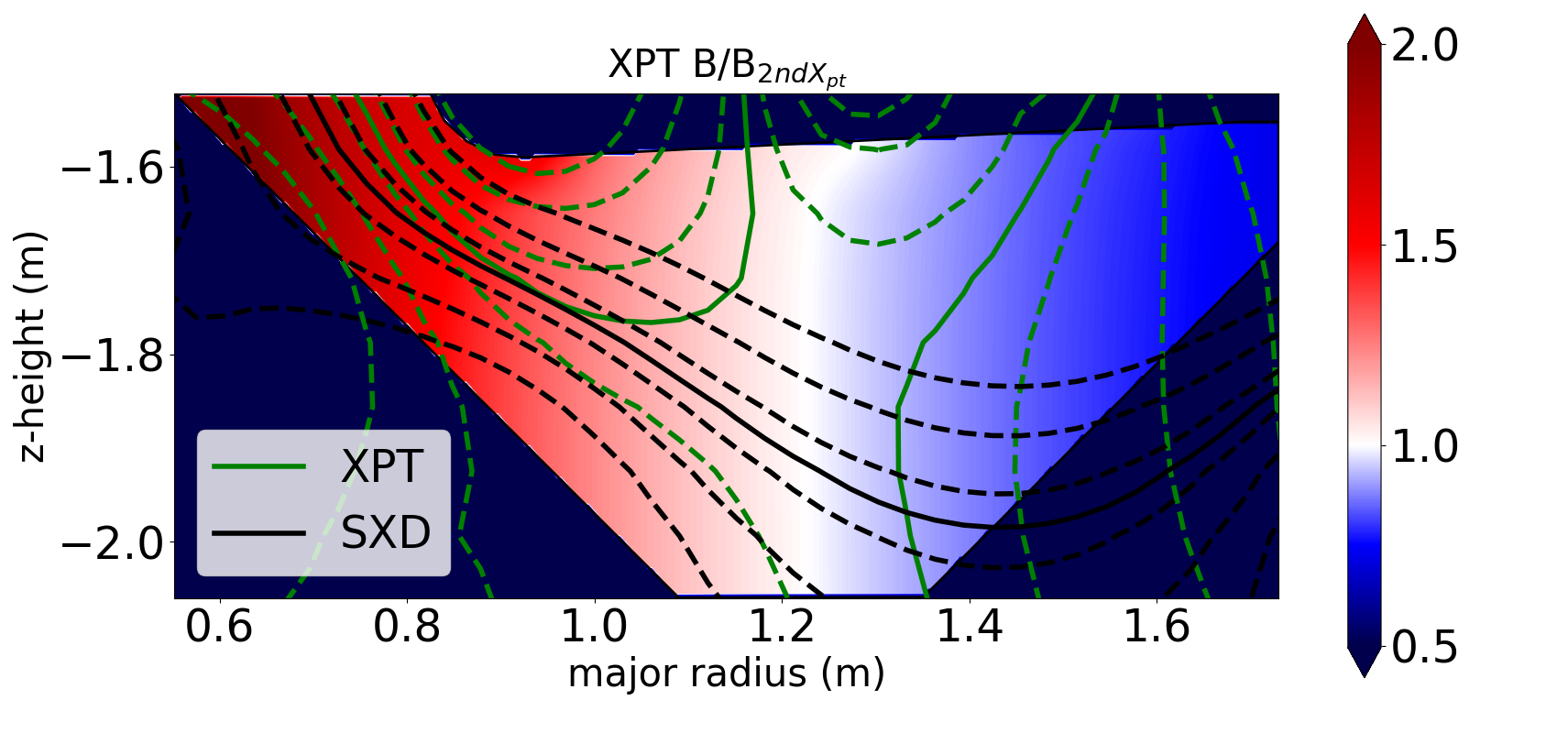}
\vspace{-0.1 cm}\caption{Ratio between the magnetic field magnitude across the divertor and the field at the 2nd X-point. Overlaid are the separatrix and flux surfaces in the SXD and XPT geometries.}
\label{fig:fig3_field}
\vspace{-0.3cm} 
\end{figure}

\begin{figure}[h!]\centering
\begin{subfigure}[b]{0.45\textwidth}
\includegraphics[ width= \textwidth]{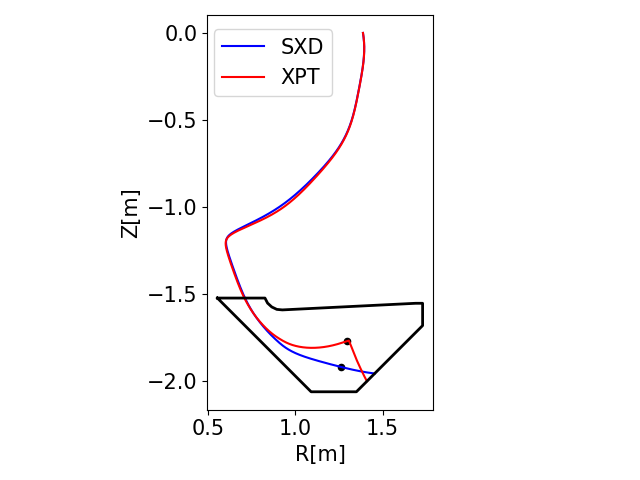} 
\caption{}
\end{subfigure}
\begin{subfigure}[b]{0.45\textwidth}
\includegraphics[width= \textwidth]{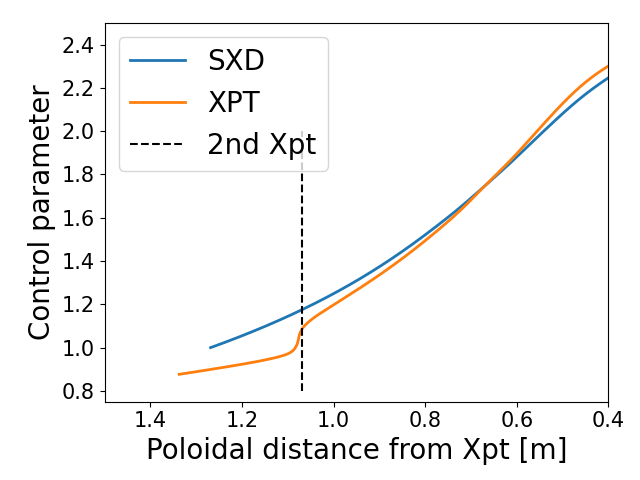}
\caption{}
\end{subfigure}
\vspace{-0.4cm} 
\caption{(a) Example flux surface at $\Bar{\psi} = 1.003$ in SXD and XPT configuration and (b) DLS prediction of detachment front location as a function of the lumped control parameter. The dashed line corresponds to the position of the 2nd null, and the corresponding points in the geometry are represented as dots in figure (a).}
\label{fig:fig4_DLS}
\vspace{-0.1cm} 
\end{figure}

\hfill \newline
According to reduced models such as the Detachment Location Sensitivity (DLS) model \cite{Lipschultz_2016}, a decreasing magnetic field magnitude along the flux tube towards the outer target should reduce the sensitivity of the detachment front location, thus slowing down the movement of the front towards the X-point. This is an advantage as it simplifies real-time control of the detachment front location \cite{kool2024demonstrationsuperxdivertorexhaust}, which is necessary to avoid the divertor re-attaching, or the detachment front reaching the primary X-point, which can result in reduced core performance or even a radiative collapse of the discharge. Due to different angles in the poloidal plane of the flux surfaces with respect to the magnetic field gradient ($\nabla|B| \sim \nabla|B_{toroidal}| \propto 1/R$, comparable in the two configurations), changes in the performance of the SXD and XPT are expected from the difference in magnetic field shape alone.  The reason for this difference is shown in figure \ref{fig:fig3_field}, in which the 2 geometries are overlaid on a 2D profile of the ratio of the magnetic field magnitude in the divertor to the magnitude at the XPT 2nd null position. In the SXD, the flux surfaces are almost parallel to the magnetic field gradient, resulting in a reduction in the sensitivity of the detachment front location from the divertor baffle entrance to the target. In the XPT, instead, the flux surfaces downstream of the 2nd null are mostly aligned with the z direction, almost perpendicular to the magnetic field gradient, and a higher sensitivity is expected in that region. \\
The two shapes can be compared quantitatively by applying the DLS model\cite{Lipschultz_2016}, which expresses the location of the front on a field-line as a function of a lumped control parameter ($C$) and the magnetic field geometry, with 
\begin{align}
    C = \frac{n_u \sqrt{f_{eff}}}{P_{div}^{5/7}}
\end{align}
and with $n_u$ the upstream density, $f_{eff}$ the effective impurity fraction and $P_{div}$ the power entering the divertor. Starting from the arbitrary $C$ corresponding to the detachment onset, increasing $C$ will move the front upstream toward the X-point with a sensitivity that depends on the magnetic geometry.
The two geometries are compared using the DLS on a field line passing close to the 2nd null in figure \ref{fig:fig4_DLS}\footnote{A slightly different XPT geometry is used for this comparison, with the 2nd null in the SOL, to ensure the DLS is applied to a flux surface reaching the outer midplane as required by the model.}. As the model is not fully predictive and focuses on the sensitivity of the front location to varying $C$ and magnetic geometry more than on absolute predictions of the location,  the values of the control parameter have been normalized to the $C$ value corresponding to the SXD detachment onset.

The main features include: 
\begin{itemize}
    \item A lower detachment threshold for the XPT, noticeable as a lower control parameter (e.g. upstream density) to move the front off the target.
    \item A small slope between the target and the secondary X-point, suggesting the detachment front will jump from the target to the 2nd null for a small variation in control parameter.
    \item A steep slope in the proximity of the 2nd X-point, suggesting a predicted reduced sensitivity of the detachment front in the proximity of the 2nd null.
    \item Comparable values once upstream of the 2nd null, in agreement with the measurements shown in figure \ref{fig:fig9_fonts}.
\end{itemize} 
The XPT's higher sensitivity near the target may have some advantages when trying to diagnose the onset of detachment, as the movement off the target should be easier to detect, but the reduced sensitivity near the secondary null may make determining the divertor state based on the detachment front position alone more difficult, possibly requiring additional diagnostic information to control the divertor state. Furthermore, the stabilization of the radiation near the secondary X-point can also behave similarly to an X-point radiator \cite{UMANSKY_XPT_ADX}\cite{Lee_XPT}, creating a condensed region of strong radiation away from the divertor walls.\\
It is worth noting that this comparison uses a 1D, single-flux tube approximation. As can be seen in figure \ref{fig:fig2_dist}, the effect of the parallel distance can vary significantly with the distance from the 2nd null. Similarly, the predicted detachment front sensitivity benefits can also change depending on which flux tube is chosen, with the most significant benefits observed closest to the 2nd null. Further analysis could compare the predicted benefits in the 1D analysis to an average of the 1D analysis across multiple flux tubes extending for a single scrape-off width.\\
\subsection{Limits in the experimental scenarios}\label{sec:XPT_caveats}
\begin{figure}[h!]\centering
\includegraphics[ width= 0.5\textwidth]{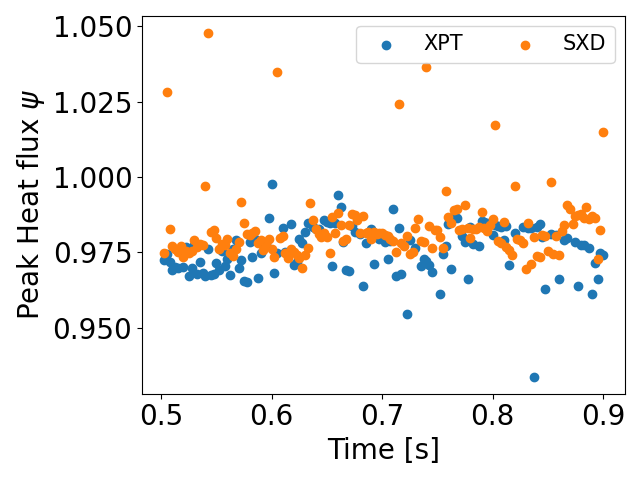} 
\vspace{-0.4cm} 
\caption{Normalized $\psi$ of the peak heat flux measured by IR cameras during the SXD and XPT discharges, highlighting how the peak is in the PFR, consistently with the $n_e$ and emission inferences. During ELMs, the peak is sometimes observed in the SOL, as can be noticed from the few dots with values greater than 1.}
\label{fig:fig011_IR_psi}
\vspace{-0.1cm} 
\end{figure}

 In previous L-mode discharges, the geometry of the outer leg was shown to have no significant effect on the core behavior for similar magnetic geometries outside of the divertor chamber\cite{verhaegh2024improved}. In these discharges, the $P_{SOL}$ is comparable, although slightly higher in the XPT discharge, while the core stored energy is lower in the XPT discharge, as shown in figure \ref{fig:fig8_overview}. The pedestal temperature is also lower in the XPT, leading to different ELM behavior, with the SXD discharge showing lower frequency type I ELMs while the XPT discharge shows type III ELMs. One potential explanation for the differences in $P_{SOL}$, stored energy, and in the pedestal temperature is the different inner strike point location, resulting in different poloidal angles between the inner divertor leg and the tile\cite{Federici_AAPPS}. An alternative explanation is the imperfect reproducibility of the H-mode scenario due to the presence of an internal reconnection event during the ramp-up and the resulting MHD mode activity. Nevertheless, the different $P_{SOL}$, stored energy, and ELM behaviour can affect the steady state performance of the divertors. The higher estimated $P_{SOL}$ in the XPT should make the XPT performance appear weaker than it would for the same $P_{SOL}$. A significant improvement is still observed, particularly at low divertor neutral pressures, implying the improvement for the same core conditions may be even stronger. The effect of the change in ELM behavior on the conclusions is less clear. The neural network results shown in section \ref{sec:vol_processes} are based on frames selected in inter-ELM periods for both configurations. The $n_e$ profiles in figure \ref{fig:fig6_CIS} are taken from an inter-ELM period for the SXD case, versus integrating over the small ELMs in the XPT case, as the CIS frame rate is too low to filter out the high-frequency ELMs in the XPT discharge. This is not expected to affect the results significantly. In the SXD discharge, both inter-ELM frames and frames that integrate over the larger type I ELMs show a comparable $n_e$ profile along the separatrix. The main difference is that the frames integrating over the ELMs show a high-density region at the target in the far SOL, where no light is present inter-ELM. This suggests that even when integrating the signal over the duration of the frame with an ELM in it, the inter-ELM contribution dominates over the ELM emission in the region of the image where there is significant emission inter-ELM. As there is significant $D_\gamma$ emission along the whole $\psi = 0.975$ flux surface inter-ELM, as observed by the MWI diagnostic operating at a higher framerate, the $n_e$ inferences near the target are unlikely to be strongly influenced. Furthermore, a significant over-estimation of $^{CIS}n_e$ near the target would result in an underestimation of $^{CIS-LP}T_e$ and an overestimation of $^{CIS-EIR}T_e$, while they are in reasonable agreement as shown in figure \ref{fig:fig12_Te}. Future work will aim for an improved matching of the shape outside the divertor chamber and for type I ELMs with similar frequency in both configurations to allow easier ELM filtering for the XPT case, as well as ensuring similar inter-ELM behaviors.\\
If the magnetic reconstruction shown in figure \ref{fig:fig01_currents} was correct, the peak heat flux would be expected on the underside of the divertor baffle, out of view of the IR camera and the Langmuir probes. As shown in figure \ref{fig:fig011_IR_psi}, however, in both SXD and XPT configurations, the heat flux inferred by the IR cameras is peaked at $\psi \approx 0.975$. 
These results are consistent with the ones in figure \ref{fig:fig5_MWI}, and \ref{fig:fig6_CIS}, showing how the Balmer emission and electron density are peaked in the PFR around $\psi \approx 0.975$. This suggests a mismatch with the EFIT magnetic reconstruction, which was accounted for by using the $\psi \approx 0.975$ surface as the reference surface for the analysis. This is particularly relevant for the XPT discharge due to the larger poloidal flux expansion near the 2nd null, increasing the distance between the $\psi = 1$ and $\psi = 0.975$ surfaces. While drifts are expected to point towards the PFR in the lower divertor, they are unlikely to be the cause of the shift of the peak loading from the separatrix towards the PFR, as a similar shift is observed in the upper divertor, where drifts point towards the far SOL. A possible explanation is the presence of plasma currents affecting the divertor geometry, which will be investigated further in future work. 
\subsection{Implications for reactors and ADC research}
The results presented in this work highlight how the divertor performance can be maximized by using multiple ADC strategies simultaneously. By combining a double-null configuration, strong baffling, total flux expansion, and the 2nd poloidal null of the X-point target, major benefits are found. This is the first experimental validation of this integrated divertor configuration, the design solution for the ARC reactor concept \cite{SPARC}, providing additional benefits over the SXD. Precise reconstructions of the 2nd null position can be very important for future devices with narrow scrape-off layers, as small shifts in the distance between the primary separatrix and the 2nd null can have strong effects on the heat flux distribution \cite{Wigram_2019_XPT}, but have proven to be challenging in this work, with more evident consequences in the XPT discharge. The presence of edge currents and toroidal field ripple, not accounted for in the EFIT reconstructions used in this work, has been proposed as a possible reasons for the mismatch and will be investigated further in future work, as accurate estimations of the 2nd null position over long time frames might be needed in future reactors relying on this configuration to ensure acceptable exhaust performance. \\ 
Furthermore, these results complement previous work showing that the benefits of total flux expansion are a continuum and both volumetric particle sinks \cite{verhaegh2024improved} and power and momentum losses \cite{lonigro20252delectrondensityprofile} continuously increase as the total flux expansion and poloidal leg length are increased. This 'mix-and-match' approach to divertor design indicates how, for given upstream conditions, a suitable exhaust solution in terms of tolerable target heat and particle fluxes can be found (without affecting the core \cite{Harrison_2024} conditions) by increasing total flux expansion, adding a secondary X-point, or a combination thereof. The divertor design of choice then depends on the magnitude of heat flux reduction needed, usually higher in compact machines such as spherical tokamaks or high-field tokamaks, which can be obtained through different combinations of these features.\\
In order to simplify the divertor design processes, however, reduced models are needed to provide a simple description of the expected benefits of these ADCs. While the DLS has proven to be a useful reduced model for comparisons of detachment access and location sensitivity, further models are needed to describe the measured power and momentum loss reduction with varying divertor geometry and through the addition of secondary X-points in the divertor. 1D models have been receiving attention for the purpose of divertor design\cite{SPLEND1D}, but no experimental validation against the XPT configuration, inherently 2D, has been performed thus far and could be the focus of future work. Similarly, 1D reduced models for the buffering of transients \cite{HENDERSON_reattachment}\cite{Flanagan_2025} could be validated against the experimental XPT data. This should be complemented by a more detailed experimental quantification of the reduction in target fluxes, both as a function of divertor conditions and with varying levels of total flux expansion. Further studies will also be needed to characterize the behaviour of the upper divertor, the sensitivity of the results on the 2nd null distance from the separatrix, and how these benefits extend to more attached and seeded conditions.
\section{Conclusions}\label{sec:conclusions}
Evidence of the improved exhaust performance of the X-point target divertor configuration on MAST-U has been presented, granting additional benefits over the Super-X divertor, which is already a major improvement over the conventional divertor.\\
The combination of the additional connection length and the wider cross-field $n_e$ profile in the proximity of the secondary X-point leads to additional plasma-neutral interactions, one of the main drivers of the benefits of the Super-X configuration after the increased interaction area with the target. These interactions increase the total hydrogenic emission in the divertor chamber, evidence of additional volumetric particle sinks, and decrease the target electron temperatures. The peak particle and heat fluxes to the target are also reduced. While the reduction in total heat flux reaching the target in the deeply detached conditions achieved on MAST-U is more modest, a significant reduction in the plasma-related heat flux is observed, suggesting favorable scaling to hotter divertor conditions. Preliminary results suggest that these benefits also extend to transients, resulting in the buffering of higher energy ELMs for the same divertor neutral pressure compared to the SXD.\\
These results add to the existing body of literature on the significant exhaust benefits that have been achieved using alternative divertor configurations, particularly in the presence of strong divertor baffling, and they highlight how multiple strategies employed in different divertor configurations can be combined to further increase their benefits.\\
While these results are promising, they are limited to strongly detached conditions due to the strong exhaust capabilities of these configurations. Future work will try to push the comparison to more attached conditions through a combination of cryopumping in the divertor and the additional NBI heating power currently being installed on MAST-U. 
\section{Acknowledgements}
This work has been carried out within the framework of the EUROfusion Consortium, partially funded by the European Union via the Euratom Research and Training Programme (Grant Agreement No 101052200 — EUROfusion), and from EPSRC Grants EP/W006839/1 and EP/S022430/1 and supported by US DOE DE-AC05-00OR22725. The Swiss contribution to this work has been funded by the Swiss State Secretariat for Education, Research and Innovation (SERI). Views and opinions expressed are however those of the author(s) only and do not necessarily reflect those of the European Union, the European Commission or SERI. Neither the European Union nor the European Commission nor SERI can be held responsible for them.

\section*{References}
\bibliographystyle{iopart-num}
\bibliography{bib}

\section*{Contributors}

\subsection*{EUROfusion Tokamak Exploitation Work Package (WPTE)}

EUROfusion Tokamak Exploitation Work Package (WPTE) Team List (2022-2023)

The EUROfusion Tokamak Exploitation Work Package (WPTE) is a Work Package established within the EUROfusion consortium to run a coordinated program across several European fusion plasma tokamak devices, namely JET, ASDEX-Upgrade, TCV, WEST and MAST-U. The WPTE Team List contain the scientific and technical team that contribute to the scientific exploitation of EUROfusion program across these devices. The EUROfusion Tokamak Exploitation Team list is updated every two years and published in the Nuclear Fusion Journal. The 2022-2023 WPTE Team list is published in Nuclear Fusion 64 (2024) 112019; DOI:10.1088/1741-4326/ad2be4 and is enclosed below

\subsection*{WPTE Team Members}

{\small
D. Abate\textsuperscript{7} , J. Adamek\textsuperscript{9} , M. Agostini\textsuperscript{7} , C. Albert\textsuperscript{10} , F. C. P. Albert Devasagayam\textsuperscript{11} , S. Aleiferis\textsuperscript{5} , E. Alessi\textsuperscript{12} , J. Alhage\textsuperscript{13} , S. Allan\textsuperscript{5} , J. Allcock\textsuperscript{5}
% ...
}

\subsection*{WPTE Affiliations}

\begin{enumerate}
  \item IRFM-CEA Centre de Cadarache 13108 Sant-Paul-lez-Durance, France
  \item Max Planck Institute for Plasma Physics, Boltzmannstrasse 2, 85748 Garching bei München, Germany
  \item ENEA, Fusion and Nuclear Safety Department, C. R. Frascati, Via E. Fermi 45, 00044, Frascati, Roma, Italy
  \item VTT Technical Research Centre of Finland, PO Box 1000, FIN-02044 VTT, Finland
  \item United Kingdom Atomic Energy Authority, Culham Science Centre, Abingdon, Oxfordshire, OX14 3DB, UK
  \item École Polytechnique Fédérale de Lausanne (EPFL), Swiss Plasma Center (SPC), CH-1015 Lausanne, Switzerland
  \item Consorzio RFX, C.so Stati Uniti 4, 35127 Padova, Italy
  \item Istituto per la Scienza e la Tecnologia dei Plasmi, CNR, Padova, Italy
  \item Institute of Plasma Physics of the CAS, Za Slovankou 1782/3, 182 00 Praha 8, Czech Republic
  \item Graz University of Technology, Petersgasse 16, 8010 Graz, Austria
  \item Aalto University, PO Box 14100, FIN-00076 Aalto, Finland
  \item Institute for Plasma Science and Technology, CNR, via R. Cozzi 53, 20125 Milano, Italy
  \item Department of Applied Physics, Ghent University, 9000 Ghent, Belgium
  \item National Technical University of Athens, Iroon Politechniou 9, 157 73 Zografou, Athens, Greece
  \item Department of Physics and Astronomy, Uppsala University, SE-75120 Uppsala, Sweden
  \item Aix-Marseille University, CNRS, PIIM, UMR 7345, 13013 Marseille, France
  \item Consorzio CREATE, Via Claudio 21, 80125 Napoli, Italy
  \item Aix Marseille Univ., CNRS, PIIM, F-13397 Marseille CEDEX 20, France
  \item Centre for Energy Research, POB 49, H-1525 Budapest, Hungary
  \item University of Milano-Bicocca, Piazza della Scienza 3, 20126 Milano, Italy
  \item Department of Electrical and Electronic Engineering, University of Cagliari, Italy
  \item Laboratorio Nacional de Fusión, CIEMAT, 28040 Madrid, Spain
  \item Electromagnetic Engineering and Fusion Science, KTH Royal Institute of Technology, SE-10044 Stockholm, Sweden
  \item Politecnico di Torino, Corso Duca degli Abruzzi, 24, 10129 Torino, Italy
  \item Dipartimento di Ingegneria Elettrica Elettronica e Informatica, Università degli Studi di Catania, 95125 Catania, Italy
  \item Institut Jean Lamour, UMR 7198, CNRS-Université de Lorraine, 54500 Vandoeuvre-lès-Nancy, France
  \item Institute of Nuclear Physics, Radzikowskiego 152, 31-342 Kraków, Poland
  \item Forschungszentrum Jülich GmbH, Institut für Energie- und Klimaforschung, Plasmaphysik, 52425 Jülich, Germany
  \item DIFFER-Dutch Institute for Fundamental Energy Research, Eindhoven, The Netherlands
  \item Institut Jean Lamour, Université de Lorraine, Nancy, France
  \item Laboratory for Plasma Physics LPP-ERM/KMS, B-1000 Brussels, Belgium
  \item Instituto de Plasmas e Fusão Nuclear, Instituto Superior Técnico, Universidade de Lisboa, 1049-001 Lisboa, Portugal
  \item Universitá degli Studi della Tuscia, DEIM Department, Viterbo, Italy
  \item KU Leuven, Department of Mechanical Engineering, Leuven, Belgium
  \item Università di Roma Tor Vergata, Via del Politecnico 1, Roma, Italy
  \item ITER Organization, Route de Vinon-sur-Verdon, CS 90 046, 13067 Saint Paul Lez Durance Cedex, France
  \item Institute of Plasma Physics and Laser Microfusion, Hery 23, 01-497 Warsaw, Poland
  \item Astrophysics Research Centre, School of Mathematics and Physics, Queen’s University, Belfast, BT7 1NN, UK
  \item Jožef Stefan Institute, Ljubljana, Slovenia--University of Ljubljana, Lyubljana, Slovenia
  \item York Plasma Institute, University of York, York, YO10 5DD, United Kingdom of Great Britain and Northern Ireland
  \item Universidad de Sevilla, Sevilla, Spain
  \item The National Institute for Laser, Plasma and Radiation Physics, Magurele-Bucharest, Romania
  \item Aix-Marseille Univ., CNRS, M2P2, Marseille, France
  \item National Science Center 'Kharkov Institute of Physics and Technology', Akademichna 1, Kharkiv 61108, Ukraine
  \item Department of Physics, Durham University, Durham, DH1 3LE, UK
  \item Department of Space, Earth and Environment, SEE, Chalmers University of Technology, SE-41296 Gothenburg, Sweden 47 Dipartimento di Ingegneria Astronautica, SAPIENZA Università di Roma, Via Eudossiana 18, 00184 Roma, Italy
  \item Barcelona Supercomputing Center, Barcelona, Spain
  \item Aix-Marseille University, CNRS, IUSTI, UMR 7343, 13013 Marseille, France
  \item Department of Physics, Technical University of Denmark, Bldg 309, DK-2800 Kgs Lyngby, Denmark
  \item Institute for Nuclear Research, Prospekt Nauky 47, Kyiv 03680, Ukraine
  \item Technische Universität Wien, Wiedner Hauptstr. 8-10/134, A-1040 Wien, Austria
  \item Laboratoire de Physique des Plasmas, Ecole Polytechnique, Palaiseau, France
  \item Institute of ion physics and applied Physics, University of Innsbruck, Austria
  \item Institute of Electronics, Bulgarian Academy of Sciences (BAS); 72 Tsarigradsko Chaussee, 1784 Sofia, Bulgaria
  \item Institute of Nuclear Physics Polish Academy of Sciences (IFJ PAN), Krakow, Poland
  \item Loughborough University, Loughborough, Leicestershire, UK
  \item Section of Astrophysics, Physics Department, Aristotle University, Thessaloniki, GR 541 24, Greece
  \item University of Helsinki, PO Box 43, FI-00014 University of Helsinki, Finland
  \item National Centre for Nuclear Research (NCBJ), 05-400 Otwock-Świerk, Poland
  \item ICREA and Barcelona Supercomputing Center, Barcelona, Spain
  \item CNR and Dipartimento di Energia—Politecnico di Torino, C.so Duca degli Abruzzi 24, 10129 Torino, Italy
  \item Universidad Carlos III de Madrid, Madria, Spain
  \item Eindhoven University of Technology, Eindhoven, The Netherlands
  \item Politecnico di Milano, Milan, Italy
  \item Rudolf Peierls Centre for Theoretical Physics, University of Oxford, Oxford OX1 3NP, United Kingdom
  \item Space and Plasma Physics, EECS, KTH SE-100 44 Stockholm, Sweden
  \item Aix Marseille Univ., CNRS, IUSTI UMR 7343, F-13013 Marseille, France
  \item EUROfusion Programme Management Unit, Boltzmannstr. 2, 85748 Garching, Germany
  \item University Roma Tre, via Vito Volterra N°62, CAP 00146 ; Rome, Italy
  \item Lithuanian Energy Institute, Laboratory of Nuclear Installation Safety, Breslaujos Str. 3, LT-44403, Kaunas, Lithuania
  \item V.N. Karazin Kharkiv National University, Kharkiv, Ukraine
  \item Center for Energy Research (CER), University of California-San Diego (UCSD), La Jolla, CA, USA
  \item Massachusetts Institute of Technology, Plasma Science and Fusion Center, Cambridge, MA 02139, USA
  \item Warsaw University of Technology, Nowowiejska 15/19, 00-665 Warsaw, Poland
  \item Faculty of Mathematics and Natural Sciences, Heinrich Heine University Düsseldorf, 40225 Düsseldorf, Germany
\end{enumerate}

\newpage
\subsection*{MAST Upgrade Team List (2022--2023)}

MAST Upgrade Team List (2022-2023)

The MAST Upgrade Team List contains the scientific and technical team that contribute to the scientific exploitation of the MAST Upgrade tokamak. The MAST Upgrade Team list is updated every two years and published in the Nuclear Fusion Journal. The 2022-2023 MAST-U Team list is published in Nuclear Fusion 64 (2024) 112017; DOI: 10.1088/1741-4326/ad6011 and is enclosed below.

\subsection*{MAST Upgrade Team Members}

{\small
J. R. Harrison\textsuperscript{1}, A. Aboutaleb\textsuperscript{2}, S. Ahmed\textsuperscript{3}, M. Aljunid\textsuperscript{1}, S. Y. Allan\textsuperscript{1}, H. Anand\textsuperscript{4}, Y. Andrew\textsuperscript{5}, L. C. Appel\textsuperscript{1}, A. Ash\textsuperscript{1}, J. Ashton\textsuperscript{1}
% ...
}

\subsection*{MAST Upgrade Affiliations}

\begin{enumerate}
  \item UKAEA (United Kingdom Atomic Energy Authority), Culham Science Centre, Abingdon, Oxfordshire, OX14 3DB, UK
  \item Department of Physics, Florida International University, 11200 SW, Miami, FL 33199, USA
  \item Department of Physics and Technology, UiT The Arctic University of Norway, N-9037 Tromsø, Norway
  \item General Atomics, PO Box 85608, San Diego, CA 92186-5608, USA
  \item Blackett Laboratory, Imperial College London - London, SW7 2BW, UK
  \item Rudolf Peierls Centre for Theoretical Physics, University of Oxford, Oxford OX1 3NP, UK
  \item Astrodel LLC, Boulder, Colorado 80303, USA
  \item Princeton Plasma Physics Laboratory, Princeton, NJ, USA
  \item Max Planck Institute for Plasma Physics, Boltzmannstrasse 2, 85748 Garching bei München, Germany
  \item Department of Electrical Engineering and Electronics, University of Liverpool, Brownlow Hill, Liverpool, L69 3GJ, UK
  \item Department of Physics and Astronomy, University of Manchester, Oxford Road, Manchester M13 9PL, UK
  \item Department of Physics, Durham University, Durham, DH1 3LE, UK
  \item Institute for Fusion Studies, The University of Texas at Austin, Austin, TX, USA
  \item York Plasma Institute, University of York, York, YO10 5DD, United Kingdom of Great Britain and Northern Ireland
  \item Physics and Astronomy Dept., University of California, Los Angeles, California 90098 USA
  \item DIFFER-Dutch Institute for Fundamental Energy Research, Eindhoven, The Netherlands
  \item Eindhoven University of Technology, Eindhoven, The Netherlands
  \item Dublin City University, Dublin, Ireland
  \item National Science Center 'Kharkov Institute of Physics and Technology', Akademichna 1, Kharkiv 61108, Ukraine
  \item École Polytechnique Fédérale de Lausanne (EPFL), Swiss Plasma Center (SPC), CH-1015 Lausanne, Switzerland
  \item Oak Ridge National Laboratory, Oak Ridge, Tennessee 37831, USA
  \item IRFM-CEA Centre de Cadarache 13108 Sant-Paul-lez-Durance, France
  \item Electromagnetic Engineering and Fusion Science, KTH Royal Institute of Technology, SE-10044 Stockholm, Sweden
  \item Department of Physics, University of Warwick, Coventry, CV4 7AL, UK
  \item Department of Physics, SUPA, University of Strathclyde, Glasgow, Scotland, UK
  \item Universidad de Sevilla, Sevilla, Spain
  \item Institute of High Performance Computing, A*STAR, Singapore
  \item Plasma Science and Fusion Center, Massachusetts Institute of Technology, Cambridge, Massachusetts 02139, USA
  \item Laboratory for Plasma Physics LPP-ERM/KMS, B-1000 Brussels, Belgium
  \item Lawrence Livermore National Laboratory, Livermore, CA, USA
  \item Dept. of Computer Science, College of William \& Mary, Williamsburg, VA, USA
  \item Department of Applied Physics and Applied Mathematics, Columbia University, New York, NY, USA
  \item Aalto University, PO Box 14100, FIN-00076 Aalto, Finland
  \item Dipartimento di Fisica "G. Galilei", Università degli Studi di Padova, Padova, Italy
  \item Consorzio RFX, C.so Stati Uniti 4, 35127 Padova, Italy
  \item ENEA, Fusion and Nuclear Safety Department, C. R. Frascati, Via E. Fermi 45, 00044, Frascati, Roma, Italy
  \item University of California Irvine, Irvine, CA 92697, USA
  \item Forschungszentrum Jülich GmbH, Institut für Energie- und Klimaforschung—Plasmaphysik, 52425 Jülich, Germany
\end{enumerate}

\end{document}

%% file: packages_commands.tex
 \usepackage{hyperref}
 \usepackage{color}
 \usepackage[utf8]{inputenc}
 \usepackage{bm}
 \usepackage{graphicx, subcaption}
 \usepackage{latexsym}
 \usepackage{pifont}
 \usepackage{amssymb}
 \usepackage{hyperref}
 \usepackage{titlesec}
 \expandafter\let\csname equation*\endcsname\relax
 \expandafter\let\csname endequation*\endcsname\relax
 \usepackage{amsmath}
 \usepackage{physics}
 \usepackage{listings}
 \usepackage{verbatim}
 \usepackage{geometry}
 \usepackage{booktabs}
 \usepackage{enumitem}
 \usepackage[normalem]{ulem} 
 \usepackage{breqn}
 \usepackage[toc,page]{appendix}
 
 \definecolor{red}{rgb}{1.0,0.0,0.0}
 \definecolor{gre}{rgb}{0.0,1.0,0.0}
 \definecolor{blu}{rgb}{0.0,0.0,1.0}

 \definecolor{ora}{rgb}{1.0,0.5,0.0}
 \definecolor{gra}{rgb}{0.0,0.5,1.0}

 \newcommand{\ba}{\begin{abstract}}
 \newcommand{\ea}{\end{abstract}}
 \newcommand{\be}{\begin{equation}}
 \newcommand{\ee}{\end{equation}}

 \newcommand{\rom}[1]{\uppercase\expandafter{\romannumeral #1 \relax}}